\documentclass[pre,aps,twocolumn]{revtex4-2}

\usepackage[english]{babel} 
\usepackage{graphicx}
\usepackage{amssymb,amsfonts,amsmath}
\usepackage{color}
\usepackage[hidelinks]{hyperref}

\usepackage{tikz}
\usetikzlibrary{shapes}
\usetikzlibrary{patterns}
\usetikzlibrary{angles, quotes}
\definecolor{bostonuniversityred}{rgb}{0.8, 0.0, 0.0}
\definecolor{chromeyellow}{rgb}{1.0, 0.65, 0.0}

\newcommand{\rv}{{\mathbf r}}
\renewcommand{\vec}{\mathbf}

\usepackage{amssymb}

\usepackage[final]{showlabels} 

\begin{document}


\title{Asymptotic methods for confined fluids}
\author{E. Di Bernardo}
\author{J.~M. Brader}
\affiliation{Department of Physics, University of Fribourg, CH-1700 Fribourg, Switzerland}


\begin{abstract}

The thermodynamics and  microstructure of confined fluids 
with small particle number are best described using 
the canonical ensemble. However, practical calculations 
can usually only be performed in the grand-canonical 
ensemble, which can introduce unphysical artifacts. 
We employ the method of asymptotics to 
transform grand-canonical observables to 
the canonical ensemble, where the former can be 
conveniently obtained using the classical density 
functional theory of inhomogeneous fluids.
By formulating the ensemble transformation as a contour integral in the complex fugacity plane we reveal the 
influence of the Yang-Lee zeros in determining the 
form and convergence properties of the asymptotic 
series. 
The theory is employed to develop expansions for the canonical partition function and the canonical 
one-body density. 
Numerical investigations are then performed using an exactly 
soluble one-dimensional model system of hard-rods.   
\end{abstract}

\maketitle

\section{Introduction}

Statistical mechanical studies of classical equilibrium 
fluids are typically 
performed using the grand-canonical ensemble, where the 
average particle number in the system is controlled by the 
chemical potential of a particle reservoir. While this 
presents several technical advantages for calculation, there 
exist many situations in which particle number fluctuations 
induce unphysical errors and a canonical treatment would be preferable. 

When can we expect differences between the grand-canonical 
and canonical ensembles to be relevant? 
The most obvious case is when the particle number is 
small, such that the system is far from the 
thermodynamic limit \cite{Hillpaper,Hillbook}. 
Small systems are naturally associated with localization of 
the particles and thus a spatially inhomogeneous 
one-body density distribution. 
In the simplest case, small groups of particles can be 
localized by the application of a confining external potential 
field, for example hard-spheres in a cavity or harmonic confining potential \cite{Evans_canonical_letter,Evans_canonical_JCP}. 
Alternatively, the localization may be a consequence of 
a mutual attraction between the particles, which binds them 
together in a finite size cluster \cite{ClusterColloid}. 
If the bulk system exhibits an equilibrium phase transition, 
e.g.~from a gas to a liquid, then the situation becomes more complicated and one has to consider more carefully how many particles constitute a `small number'. 
If the thermodynamic parameters are close to those of 
the bulk critical point, then even a confined fluid with 
many particles can exhibit finite-size effects
due to the development of long-ranged correlations.

Localized systems with small particle number have a  
microstructure and thermodynamics which differs 
significantly from those of an infinite bulk 
system; interfacial effects become dominant. 
The most commonly employed and well studied framework 
for determining the 
equilibrium properties of inhomogeneous classical fluids is the 
Density Functional Theory (DFT) \cite{Evans79,Evans92}.
This approach has proved enormously useful in the study of 
inhomogeneous fluids but exhibits grand-canonical artifacts in situations for which 
the particles are strongly localized 
\cite{Evans_canonical_letter,Evans_canonical_JCP,
ExactCanonical,ReinhardtBraderLocalized}. 
Although DFT has been formally recast in the canonical ensemble \cite{white_canonical,white_equiv}, 
the complexity of implementation has hindered its application to concrete problems. 
In the rare cases for which 
grand-canonical quantities are known exactly a matrix inversion scheme may be employed to obtain exact canonical 
information \cite{ExactCanonical}.
However, when the grand-canonical input is only approximately 
known, as is usually the case for models of interest, the method becomes unstable as errors in the grand-canonical input quantities become amplified. 

An alternative, albeit approximate, route to obtaining canonical quantities was investigated by Salaceuse 
{\it et al.} \cite{salacuse}, 
who expanded the canonical structure factor as a power series 
in $N^{-1}$, aiming to account for leading-order finite-size effects in simulation data. 
This approach, which builds upon early work of Lebowitz and Percus \cite{Lebowitz}, was subsequently extended to 
inhomogeneous systems by 
Gonz\'alez {\it et al.} to approximate the canonical one-body 
density profile 
\cite{Evans_canonical_letter,Evans_canonical_JCP}. 
A second-order truncation of their expansion was shown 
to yield very good results for the density 
profile of hard-spheres in a hard spherical cavity. 
The key to the success of these methods lies in the rapid convergence of the partial sums of the series, since calculation of the 
higher-order coefficients soon becomes computationally demanding. 

In this paper the transformation of observables from the grand-canonical to the canonical ensemble is formulated as a contour integral in the complex fugacity plane. 
From this starting point we use the method of steepest 
descents \cite{dingle,olver,bender_orszag} to develop asymptotic expansions for both the canonical 
partition function and canonical one-body density profile, 
although extension to higher-order correlation functions 
would not pose any additional difficulties.
Typical applications of steepest descents in the 
mathematical literature concern the approximate 
evaluation of intractable integrals for large values 
of a control parameter. 
Although the steps involved in constructing valid 
asymptotic series surely require much thought and 
subtlety, one at least knows the integrand as a 
closed-form analytic expression over the entire integration domain.
When seeking to apply asymptotics to transform between 
ensembles we do not have this luxury. 
Grand-canonical observables can usually only be calculated 
(numerically) for positive real values of the fugacity, whereas 
the contour integral for transforming to the canonical 
ensemble demands grand-canonical input calculated over 
the full complex fugacity plane. 
This limited knowledge of the integrand hinders determination of the optimal 
steepest descents contour. 
Careful consideration and ingenuity are thus required to find a contour capable 
of generating correctly the leading-order terms in the 
series, with the hope that fast convergence makes 
calculation of higher-order terms unneccessary.

Although the asymptotic method is approximate it can 
sometimes offer deeper insight into the essential elements 
of a physical problem than would be available from a 
complicated exact solution. 
In the present case we find that the asymptotic 
expansion of canonical quantities is intimately connected 
with the presence of singular points of the grand partition 
function in the complex fugacity plane: the famous 
Yang-Lee zeros. 
These nontrivial singularities play an important role in 
determining the optimal steepest-descents integration 
contour and therefore affect the convergence properties of the asymptotic series. 
Viewing the problem of ensemble transformation through the lens of asymptotics not only suggests new lines of attack 
but also sheds light on earlier studies in which the 
status of the approximations employed was not entirely 
clear.
For example, asymptotics reveals the Yang-Lee related 
pitfalls which would surely be encountered when attempting 
to apply the expansion of Gonz\'alez \textit{et al.} to realistic model fluids exhibiting a phase transition. 
The approach developed here thus brings together in 
an intuitive way the fields of classical DFT, ensemble transformation and the Yang-Lee theory of phase transitions.

\section{Density functional theory}\label{DFT}

The primary method for obtaining grand-canonical information 
about confined or localized systems is the classical 
density functional theory (DFT). 
The central object in DFT is the grand potential functional \cite{Evans79,Evans92}
\begin{align}\label{omega}
\Omega[\rho] = \mathcal{F}[\rho]  
- \int \!d\rv \big( \mu - V_{\text{ext}}(\rv) \big)\rho(\rv), 
\end{align}
where $\mu$ is the chemical potential and $V_{\text{ext}}(\rv)$ is the external potential. 
The Helmholtz free energy, 
$\mathcal{F}$, appearing on the right hand-side of equation \eqref{omega} is here a grand-canonical quantity and not the true canonical Helmholtz free energy for a system of fixed particle number.
The ideal gas and interaction contributions to 
$\mathcal{F}$ can be separated according to 
$\mathcal{F}\!=\!\mathcal{F}^{\,\text{id}} 
+ \mathcal{F}^{\,\text{exc}}$.  
The ideal gas term is given exactly by	
\begin{align}
\mathcal{F}^{\,\text{id}}[\rho]=k_BT\int\!d\rv\, \rho(\rv)\big(\ln(\rho(\rv))-1\big),
\end{align}
where $k_BT$ is the thermal energy and where the (physically irrelevant) thermal wavelength has been set equal to unity. 
The excess Helmholtz free energy functional, 
$\mathcal{F}^{\,\text{exc}}
$, encodes the interparticle 
interactions and usually has to be approximated.
The grand potential satisfies the variational condition
\begin{align}
\label{variation omega}
{\frac{\delta  \Omega[\rho\,]}{\delta \rho(\rv)}}\bigg\rvert_{\rho_{\text{eq}}}=0,
\end{align}
which then generates the Euler-Lagrange equation 
for the equilibrium density
\begin{align} \label{EL}
\rho_{\text{eq}}(\rv)=\rm{e}^{ -\beta\left(V_{\text{ext}}(\rv) - \mu - k_BT \, c^{(1)}(\rv; [\rho_{\text{eq}}])\right)}.
\end{align}
The one-body direct correlation function is a functional 
of the one-body density and is defined by the derivative
\begin{align} \label{c1 definition}
c^{(1)}(\rv; [\rho_{\text{eq}}]) = - {\frac{\delta \beta F^{\,\text{exc}}[\rho]}{\delta\rho(\rv)}}\bigg\rvert_{\rho_{\text{eq}}}.
\end{align}
Substitution of the equilibrium density profile into 
the functional \eqref{omega} generates the thermodynamic 
grand potential, 
$\Omega_{\text{eq}}\!=\!\Omega[\rho_{\text{eq}}]$, 
which is related to the grand partition function, $\Xi$, 
via the standard relation $\Omega_{\text{eq}}\!=\!-k_BT\ln(\Xi)$.  
Note that we will henceforth drop the subscript, 
$\rho_{\text{eq}}\equiv \rho$, since we will deal 
solely with equilibrium systems.

\section{Canonical partition functions and thermodynamic quantities}\label{partition}

\subsection{Exact forwards and backwards transforms}\label{forwardsBackwardsPartition}

The grand-canonical partition function is defined as a 
weighted sum of the canonical partition functions, $Z_N$, 
according to
\begin{equation}\label{forwards}
\Xi(\lambda) 
=
1 + \sum_{N=1}^{\infty} \lambda^N Z_N \,,
\end{equation}
where the fugacity is related to the chemical potential by $\lambda\!=\!\exp(\beta\mu)$. 
If the number of particles which the system can accommodate 
is limited by the physical situation under consideration,  
e.g.~particles confined within a closed cavity,  
then the sum in \eqref{forwards} can be truncated at a maximum value, 
$N\!=\!N_{\text{max}}$. 
The relation \eqref{forwards} is essentially a discrete 
Laplace  transform of the canonical partition function 
(depending on the discrete index $N$) to the 
grand-canonical partition function (a function of the 
continuous variable $\mu$). 
We will henceforth refer to this as the `forward' transform.  

The grand-canonical partition function contains 
all statistically relevant information about the equilibium 
system. 
It should therefore be possible in principle to invert the transformation \eqref{forwards} to obtain $Z_N$ from 
the grand partition function. 
In reference \cite{ExactCanonical} it is shown 
that if $\Xi$ is known at $N_{\text{max}}$ distinct trial values 
of the fugacity, $\lambda_1\cdots\lambda_{N_{\text{max}}}$, 
then equation \eqref{forwards} generates a set of algebraic equations 
for the unknown canonical partition functions, 
$Z_1\cdots Z_{N_{\text{max}}}$. 
If the grand partition function is exact, then the canonical partion functions resulting 
from solution of this system of equations are also exact, 
with values independent of the chosen trial fugacities. 
Unfortunately, this scheme is of limited applicability, since we are usually forced to 
employ approximations to the grand potential (and hence the partition function). 
This renders the algebraic inversion of \eqref{forwards} 
ambiguous, as the values obtained for the canonical 
partition functions become dependent upon the 
arbitrary choice of trial fugacities.  
The instability of direct numerical inversion of \eqref{forwards} is perhaps not surprising, given 
experience with established numerical Laplace inversion 
schemes (see e.g.~\cite{cohen}). 
It is thus worth searching for alternatives which are 
more robust for applications and which may provide a 
deeper level of insight.

Although only positive real values of $\lambda$ 
yield a physically meaningful grand partition function, 
the polynomial sum in \eqref{forwards} remains a well-defined mathematical quantity 
for complex values of $\lambda$. 
The exact inverse (back) transform corresponding to equation 
\eqref{forwards} has appeared sporadically in the statistical mechanics literature at various points over the last century (see, e.g.~\cite{fowler_book,kubo,Lebowitz,schroedinger})
and requires knowledge of 
$\Xi$ over the full complex fugacity plane. 
The back transform is given by the following contour integral
\begin{equation}\label{backwards}
Z_N = \frac{1}{2\pi i}
\oint\limits_{C} d\lambda\, \frac{\Xi(\lambda)}{\lambda^{N+1}}
\,,
\end{equation}
where the contour $C$ can be chosen arbitrarily, provided that it encloses the pole at 
$\lambda\!=\!0$. We note, however, that the contour-independence of $Z_N$ is only guaranteed when $\Xi$ is known exactly.  
The most familiar occurance in statistical mechanics of an integral of this type is as part of the Darwin-Fowler derivation of the canonical ensemble \cite{huang,fowler_book}. 
A short explanation of equation \eqref{backwards} is given in appendix \ref{proof_integral}.

The easiest way to evaluate 
the back-transform \eqref{backwards} would appear to be direct 
integration around some convenient contour enclosing the 
origin. 
The simplest choice we can make is a circular path, 
parameterized according to $\lambda\!=\!r\exp(i\phi)$, where 
the value of $r$ should be irrelevant for the outcome.  
The back-transform then reduces to
\begin{align}\label{explicit_back}
Z_N = \frac{1}{2\pi r^N}
\!\int_0^{2\pi}\!
d\phi \,
\Big(\,
&\Xi_R(r,\phi)\cos(N\phi)
\notag\\
+\,
&\Xi_I(r,\phi)\sin(N\phi)
\Big),
\end{align}
where $\Xi_R$ and $\Xi_I$ are the real and imaginary parts 
of the grand partition function, respectively, and where 
we have used the fact that $Z_N$ is real.
Given some expression for $\Xi$, which can be either 
exact or appoximate, equation \eqref{explicit_back} 
then generates a prediction for $Z_N$ for any integer value 
of the particle number $N$. 
The catch, however, is that implementing this procedure requires knowledge of 
the complex-valued grand partition function for all 
values of $\lambda$ around a circle of radius $r$ 
in the complex plane; information which is typically not available. 
In practice we usually only have access to approximate information about $\Xi$ on the positive real axis of the complex $\lambda$ plane. 
A systematic way to deal with this difficulty, 
and to hopefully get the most out of the limited information available, 
is to use the method of asymptotics to approximate the integral in \eqref{backwards} as a power series in $N^{-1}$. 
Although the asymptotic method is not exact it provides 
a route to unambiguous calculation of the canonical partition functions, even in cases for which $\Xi$ 
is not known exactly.

\subsection{Steepest-descents approximation}
\label{steepest}

\subsubsection{General considerations}

Using the Cartesian representation of the complex fugacity, 
$\lambda\!=\!x+iy$, 
we begin by considering the behavior of the integrand in equation 
\eqref{backwards}, namely 
\begin{equation}
    I(\lambda)=\frac{\Xi(\lambda)}{\lambda^{N+1}},
\end{equation}
along the positive real axis, $\lambda\!=\!x$. 
The definition \eqref{forwards} and the positivity of 
$Z_N$ imply that 
$\Xi(x)$ is positive for $x\!>\!0$. 
The thermodynamic relation 
$\langle N\rangle\!=\!-\partial\,\Omega/\partial\mu$, 
where $\langle N\rangle$ is the average number of particles in the grand-canonical system, can be expressed in the alternative form 
\begin{equation}\label{Nbar_definition}
\langle N\rangle=\frac{x}{\Xi}\frac{\partial\,\Xi}{\partial x}.
\end{equation}
Since for positive values of $x$ both $\langle N\rangle$ and $\Xi$ are positive, it follows that the derivative 
$\partial \Xi/\partial x\!>\!0$.   
The grand partition function is thus a monotonically 
increasing function of $x$. 
As the function $1/x^{N+1}$ is a monotonically decreasing 
function of $x$ we can conclude that the integrand $I(x)$ has a minimum, which we denote by $x_0$, on the 
positive real axis. 
Both  $\Xi$ and  $1/x^{N+1}$ are analytic functions of the fugacity; 
recall that $\Xi$ is simply an integer-power polynomial 
in $\lambda$. 
Consequently, the integrand $I(\lambda)$ obeys the 
second-order Cauchy-Riemann condition \cite{boas}
\begin{equation}\label{cauchy_riemann}
    \frac{\partial^2 I}{\partial x^2}+\frac{\partial^2 I}{\partial y^2} =0.
\end{equation}
Since $x_0$ is a minimum along the real axis, it follows that 
\begin{equation}
\frac{\partial^2 I}{\partial x^2}\bigg|_{{
\substack{x\,=\,x_0\\\!\!\!y\,=\,0}}}
>\,0 \,,
\qquad
\frac{\partial^2 I}{\partial y^2}\bigg|_{{
\substack{x\,=\,x_0\\\!\!\!y\,=\,0}}}
<\,0 \,,
\end{equation}
namely that  
there is a maximum of $I(\lambda)$ in 
the $y$-direction, parallel to the imaginary axis, 
at the point $\lambda\!=\!x_0$.  
We have thus identified a saddle point at $\lambda\!=\!x_0$. 

Using a simple rearrangement, equation \eqref{backwards} can 
be put into the following standard form
\begin{equation}\label{laplace_form}
Z_N = \frac{1}{2\pi i}
\oint\limits_{C} d\lambda\, \frac{e^{N f(\lambda)}}{\lambda}
\,,
\end{equation}
where we have defined 
\begin{equation}\label{f_function}
f(\lambda) = \frac{1}{N}\ln \Xi(\lambda)\; - \; \ln \lambda,   
\end{equation}
which is the negative of the reduced Helmholtz free energy 
per particle, namely 
\begin{equation}\label{reduced_helmholtz}
f=-\frac{\beta \mathcal{F}}{N}. 
\end{equation}
As previously mentioned, the Helmholtz free energy 
appearing in \eqref{reduced_helmholtz} is a grand-canonical 
quantity. 
By now treating $N$ as a large parameter we can apply the method of steepest descents \cite{bender_orszag,dingle,olver,morse_feshbach,matthews_walker} to equation \eqref{laplace_form} and thus develop an 
asymptotic expansion of the canonical partition function 
in powers of $N^{-1}$.  
The saddle-point, $x_0$, about which we wish to expand is easily located by solution of the equation $df(x)/dx\!=\!0$, where the derivative is taken along the real axis. This leads to the saddle condition
\begin{align}\label{saddle_condition}
\left(x\frac{\partial}{\partial x}\Xi(\lambda)\right)
\bigg|_{{
\substack{x\,=\,x_0\\\!\!\!y\,=\,0}}} 
= N\,\Xi(x_0). 
\end{align}
Combining equation \eqref{saddle_condition} with the relation \eqref{Nbar_definition} shows 
that the saddle condition implies
\begin{equation}
\left.\langle N\rangle\right|_{x_0}\!=\!N, 
\end{equation}
namely that the fugacity at the 
saddle-point is that which sets the 
average number of particles in the grand-canonical system equal to the sharp particle number in the canonical system.

\subsubsection{The optimal choice of contour}
\label{optimalcontour}

The next question concerns the best choice of contour passing through the saddle point. 
Since the only formal requirement of the back transform 
\eqref{backwards} is that the contour 
encloses the origin we have some freedom to optimize our 
choice to take best advantage of the available information, 
i.e.~the values of $\Xi$ along the positive real axis. 
In the following we assume that the function $f$ is analytic, which will turn out to be true, except at isolated points.

Separating $f$ into its real and imaginary 
parts, $f\!\equiv\!f_\text{re} + i f_\text{im}$,  
the directional derivatives in a direction given by a unit vector 
$\vec{n}$ are 
\begin{align}
\frac{d f_\text{re}}{ds}&=\vec{n}\cdot\nabla f_\text{re},
\label{directionalA}
\\
\frac{d f_\text{im}}{ds}&=\vec{n}\cdot\nabla f_\text{im},
\label{directionalB}
\end{align}
where $ds$ is a line element along $\vec{n}$.
If we make the specific choice 
$\vec{n}\!=\!\nabla f_\text{re}/|\nabla f_\text{re}|$, 
then we obtain the largest possible value for the right-hand side of \eqref{directionalA}, namely
\begin{align}\label{directional2}
\frac{d f_\text{re}}{ds}=|\nabla f_\text{re}|.
\end{align}
from which we conclude that 
$f_\text{re}$ changes most rapidly 
in the direction of $\nabla f_\text{re}$. 
Equation \eqref{directionalB} becomes
\begin{align}\label{directional3}
\frac{d f_\text{im}}{ds} 
=
\frac{\nabla f_\text{re}\cdot\nabla f_\text{im}}{
|\nabla f_\text{re}|}.
\end{align}
The two Cauchy-Riemann conditions  
$\partial f_\text{re}/ \partial x
\!=\! \partial f_\text{im}/ \partial y$ and 
$\partial f_\text{re}/ \partial y
\!=\! -\partial f_\text{im}/ \partial x$ \cite{boas} 
can be combined into the single expression
\begin{equation}
\left(\frac{\partial f_\text{re}}{\partial x}\right)
\left(\frac{\partial f_\text{im}}{\partial x}\right)
+ 
\left(\frac{\partial f_\text{re}}{\partial y}\right)
\left(\frac{\partial f_\text{im}}{\partial y}\right)
=0,
\end{equation}
which is simply the vectorial condition 
$\nabla f_\text{re}\cdot \nabla f_\text{im}\!=\!0$. 
Using this result in equation \eqref{directional3} leads to 
\begin{equation}
\frac{d f_\text{im}}{ds}=0.
\end{equation}
We can thus conclude that along a contour with 
tangent parallel to $\nabla f_\text{re}$ the function $f$ 
will have a constant imaginary part and a real part which 
exhibits the most rapid possible rate of change.

The above observations are very useful in evaluating 
\eqref{laplace_form}. Deforming the contour into a 
path along which the imaginary part of the function 
$f\!\equiv\!f_\text{re} + i f_\text{im}$ remains constant 
enables us to factor out a complex exponential 
which would otherwise generate inconvenient 
oscillations in the integrand. 
This yields
\begin{equation}\label{factored}
Z_N = \frac{e^{iN f_\text{im}}}{2\pi i}
\oint\limits_{\mathcal{C}^*} d\lambda\, 
\frac{e^{N f_\text{re}(\lambda)}}{\lambda}
\,,
\end{equation}
where $\mathcal{C}^*$ indicates that we are integrating along 
a contour of constant $f_\text{im}$. 
Furthermore, if we demand that $\mathcal{C}^*$ passes through the saddle point on the real axis, where $f_\text{im}\!=\!0$, it follows that $f_\text{im}$ 
is not simply constant, but equal to zero around the entire contour. 
Choosing the contour $\mathcal{C}^*$ ensures that the 
real part of the integrand decreases as rapidly as possible 
from its value at the saddle point (`steepest descent') 
\cite{bender_orszag}. 
The challenge when implementing the steepest descents scheme is that determination of the optimal contour $\mathcal{C}^*$ requires knowledge of 
$f$, and thus $\Xi$, over the entire complex $\lambda$ 
plane, whereas standard statistical mechanical methods (such as DFT) can only provide information along the real axis. 
In the following we will show how to deal with 
this difficulty.

Without loss of generality we begin by expressing the complex fugacity $\lambda$ in polar form 
\begin{equation}\label{phaseform}
\lambda = r e^{i\phi},
\end{equation}
and then substitute this into the expression 
\eqref{forwards} to obtain
\begin{eqnarray}\label{partition_polar}
\Xi_{\phi}(r) = 
1 + Z_1 \,re^{i\phi} + Z_2 \,r^2 e^{2i\phi} 
+ Z_3 \,r^3 e^{3i\phi} + \cdots\,,\,
\end{eqnarray}
where we have introduced the convenient shorthand notation 
$\Xi_{\phi}(r)\!=\!\Xi(r,\phi)$.
Taylor expansion of the exponential functions and collecting powers of $\phi$ then yields
\begin{align}\label{expanded}
\Xi_{\phi}(r) &= 
\;(1 + r Z_1 + r^2 Z_2 + r^3 Z_3 + \cdots)
\notag\\
&+ i\phi \,(r Z_1 + 2 r^2 Z_2 + 3 r^3 Z_3 + \cdots)
\notag\\
&+ \frac{(i\phi)^2}{2} \,(r Z_1 + 4 r^2 Z_2 + 9 r^3 Z_3 + \cdots)
\notag\\
&+ \frac{(i\phi)^3}{6} \,(r Z_1 + 8 r^2 Z_2 + 27 r^3 Z_3 + \cdots)
\notag\\
&+\cdots\,.
\end{align}
Starting from equation \eqref{partition_polar} we observe that repeated 
application of the operator 
$r\frac{\partial}{\partial r}$
to the grand-canonical partition function (at zero phase-angle) generates
\begin{equation}\label{spot_the_pattern}
\left(r\frac{\partial}{\partial r}
\right)^{\!n}\!\Xi_0(r)
\,=\,
\sum_{j=1}^{\infty}\;
j^{n} r^{j} \;Z_j \,.
\end{equation}
For $n\!=\!1$ equation \eqref{spot_the_pattern} generates 
the coefficient of $i\phi$ in equation \eqref{expanded}, 
$n\!=\!2$ generates the coefficient of 
$(i\phi)^2/2$ and so on. 
This enables us to write equation \eqref{expanded} in 
the more compact form 
\begin{align}\label{almost_there}
\Xi_{\phi}(r) &= 
\Xi_0(r)
+ i\phi \left( r\frac{\partial}{\partial r}\right)\Xi_0(r)
+ \frac{(i\phi)^2}{2}\! \left( r\frac{\partial}{\partial r}\right)^{\!\!2} \Xi_0(r)
\notag\\
&+ \frac{(i\phi)^3}{6}\! \left( r\frac{\partial}{\partial r}\right)^{\!\!3}\Xi_0(r)
+\cdots\,,
\end{align}
which can be formally resummed to obtain 
\begin{equation}\label{final_form}
\Xi_{\phi}(r) = 
\left(e^{
i\phi r\frac{\partial}{\partial r}
}
\right)\Xi_0(r).
\end{equation}
Equation \eqref{final_form} seems to us to be an 
interesting and general result, the derivation of which relies solely on  
the polynomial form of the grand partition function. 
It tells us that if we know $\Xi_0(r)$ and its derivatives at a point 
$r$ on the positive real axis ($\phi\!=\!0$), 
then application of the exponential `rotation operator' 
will generate the value of the grand partition function at any point on a circle of radius $r$ in the complex $\lambda$ plane. 

Let us now investigate the terms appearing in the Taylor expansion \eqref{almost_there}. From the definition of the grand partition function it is straightforward 
to show that
\begin{align}\label{eigenvalue}
\left(i\phi r\frac{\partial}{\partial r}\right)^{\!\!n}
\!\Xi_0(r) 
= i\phi \langle N^n \rangle\,\Xi_0(r),  
\end{align}
for $n\!=\!1,\cdots,\infty$.
We recall that for $\phi\!=\!0$ the radial coordinate 
$r$ is simply the fugacity on the positive real axis and 
that $\langle\cdot\rangle$ indicates a grand-canonical average quantity calculated at this value of $r$. 
The Taylor series \eqref{almost_there} can therefore be 
reexpressed as follows
\begin{align}\label{new_taylor}
\Xi_{\phi}(r) \!=\! 
\bigg(
\!1 + i\phi \langle N \rangle 
+ \!\frac{(i\phi)^2}{2}\! \langle N^2 \rangle
+ \!\frac{(i\phi)^3}{6}\! \langle N^3 \rangle 
+\cdots \!\!\bigg)
\Xi_0(r)
.
\end{align}
The contour we seek could in principle 
be determined using the following procedure: 
(i) Substitute the expansion 
\eqref{new_taylor} and polar form \eqref{phaseform} into the definition of $f$ to obtain
\begin{equation}\label{f_reexpressed}
f_{\phi}(r) = \frac{1}{N}\ln\left(
\Xi_{\phi}(r) 
\right)
- \ln\left(
r e^{i\phi}
\right),
\end{equation}
where we follow the notation already adopted for the grand partion function.
(ii) Set the imaginary part of $f_{\phi}(r)$ equal to zero 
and solve the resulting equation to find the function 
$r(\phi)$ which maps out the contour $\mathcal{C}^*$. 
Unfortunately, practical implementation of this scheme 
proves both cumbersome and numerically delicate. 
When using DFT, $\Xi_0(r)$ is generated 
using an iterative numerical procedure and the finite-difference derivatives required to calculate 
the moments $\langle N^n \rangle$ decrease in accuracy 
as the value of $n$ increases.

For these reasons we look instead for an {\it approximate} 
contour which is more convenient for practical calculations, 
but still sufficient to capture correctly the low-order 
terms in the asymptotic expansion. 
The simplest way to achieve this is to neglect particle number 
fluctuations by making 
the factorization approximation 
$\langle N^n \rangle\!\approx \langle N \rangle^n$.
Within this approximation equation \eqref{new_taylor} 
becomes
\begin{align}\label{factor_series}
\Xi_{\phi}(r) \approx 
\bigg(
\!1 + i\phi \langle N \rangle  
+ \!\frac{(i\phi \langle N \rangle)^2}{2}\!  
+ \!\frac{(i\phi \langle N \rangle)^3}{6}\!    
+\cdots \!\!\bigg)
\Xi_0(r).
\end{align}
which can then easily be resummed to give the following simple formula
\begin{align}\label{final_approx}
\Xi_{\phi}(r) \approx 
e^{i\phi \langle N\rangle}\,\Xi_0(r).
\end{align}
Substitution of \eqref{final_approx} into 
\eqref{f_reexpressed} then gives
\begin{equation}\label{approx_f}
f_{\phi}(r) \approx
\left(\frac{1}{N}\ln\left( \Xi_0(r) \right) 
- \ln\left( r\right)\right) 
+
i\phi\left( \frac{\langle N\rangle}{N} - 1 \right), 
\end{equation}
from which we identify a clear separation into real and imaginary parts. 
If we now choose the fugacity to be equal to that at the saddle point, $r\!=\!x_0$, then $\langle N \rangle\!=\!N$ and the imaginary part of \eqref{approx_f} vanishes, 
leaving us with  
\begin{equation}\label{approx_f}
f_{\phi}(x_0) \approx
\frac{1}{N}\ln\left( \Xi_0(x_0) \right) 
- \ln\left( x_0\right) , 
\end{equation}
which is a real function.
We have thus demonstrated that 
{\it within the factorization approximation} 
a circular contour, 
parameterized by the phase-angle and passing through the saddle 
point, leads to a constant imaginary 
part of $f$ and that the value of this constant is zero. 

If we would not make the factorization approximation and use 
the exact equation \eqref{new_taylor}, rather than equation \eqref{final_approx}, then the true $\mathcal{C}^*$ would be given 
by a non-circular (and nontrivial) path through the saddle 
point. 
For small values of $\phi$ the leading terms in \eqref{new_taylor} will dominate and the factorization approximation \eqref{factor_series} should be reliable.  
The circle will then provide a good approximation to 
$\mathcal{C}^*$ in the vicinity of the saddle.   
For larger values of $\phi$  deviations can be anticipated 
as the fluctuations terms we have neglected, which at 
order $i$ in the phase angle expansion \eqref{new_taylor} 
are proportional to $\langle N^i\rangle-\langle N\rangle^i$, 
become important. 
Moreover, the factorization approximation can be expected to break down close to any singular points in $f$ which
are intimately related to the growth of fluctuations. 
We will return to this point in detail in subsection 
\ref{YangLee}, where 
we employ an exactly soluble model as a numerical testbed.

\subsection{Asymptotic expansion of the canonical\\ 
partition function}\label{asymptotic_partition}

We now employ the steepest descents method to develop an 
asymptotic expansion for the canonical partition function. 
In all calculations from here on we will employ the circle approximation to the contour $\mathcal{C}^*$. 
Along this contour we can rewrite 
equation \eqref{laplace_form} as an
integral over the phase-angle
\begin{equation}
Z_N = \frac{1}{2\pi x_0}\int_{-\pi}^{\pi} e^{Nf(x_0e^{i\phi})}d\phi.
\end{equation}
Taylor expanding $f$ around the saddle, $\phi\!=\!0$, gives

\begin{equation}\label{asymptotic_start}
Z_N=\frac{e^{Nf_0}}{2\pi}
\!\int_{-\pi}^\pi \!
d\varphi\, e^{N\left( \frac{1}{2!}x_0^2(e^{i\varphi}-1)^2f_2
\;+\;
\frac{1}{3!}x_0^3 (e^{i\varphi}-1)^3f_3
\;+\;
\dots 
\right)},
\end{equation}
where we have used the shorthand notation 
\begin{equation}
\frac{\partial f}{\partial x}
= f^{(i)}(x_0)
=f_i, 
\end{equation}
and where we note that $f_1=0$, since $x_0$ is a saddle point. 
We now Taylor expand the factors $(e^{i\varphi}-1)$ in powers of $\varphi$ to obtain the following expression
\begin{align}\label{partial_expansion}
Z_N=\frac{e^{Nf_0}}{2\pi}\int_{-\pi}^\pi d\varphi 
e^{-\frac{1}{2}\kappa\varphi^2}\bigg(e^{iA\varphi^3+B\varphi^4+iC\varphi^5+D\varphi^6\cdots}
\bigg).
\end{align}
The first few
coefficients appearing in the exponentials are given by
\begin{align}\label{asymptotic_coefficients}
\kappa&=N\Lambda_2
\\
A&=N\!\left(-\frac{1}{2}\Lambda_2-\frac{1}{6}\Lambda_3\right)
\notag\\
B&=N\!\left(\frac{7}{24}\Lambda_2+\frac{1}{4}\Lambda_3+\frac{1}{24}\Lambda_4\right)
\notag\\
C&=N\!\left(\frac{1}{8}\Lambda_2+\frac{5}{24}\Lambda_3+\frac{1}{12}\Lambda_4+\frac{1}{120}\Lambda_5\right)
\notag\\
D&=N\!\left(-\frac{31}{720}\Lambda_2-\frac{1}{8}\Lambda_3-\frac{13}{144}\Lambda_4-\frac{1}{48}\Lambda_5-\frac{1}{720}\Lambda_6\right), 
\notag
\end{align}
where $\Lambda_i=x_0^if^{(i)}(x_0)$. 
Finally, we Taylor expand the bracketed factor in the integrand of \eqref{partial_expansion} to obtain a sum 
of simple integrals. 
Assuming that the integrand of each term decreases rapidly to zero as $\phi$ 
is increased away from $\varphi\!=\!0$ 
we extend the limits of integration 
and compute the resulting Gaussian integrals one by one.  
Observing that the coefficients 
\eqref{asymptotic_coefficients} all scale linearly in $N$ and that the Gaussian integrals 
scale as 
\begin{equation}
\int_{-\infty}^\infty d\varphi\, \varphi^j \,e^{-\frac{1}{2}\kappa\varphi^2}
\sim
N^{-\frac{j}{2}}
\end{equation}
for even values of the index $j$, we group the terms 
according to their dependence on $N$ and thus obtain 
our final expression for the asymptotic expansion  
\begin{equation}\label{final_expansion_ZN}
\boxed{
Z_N=\frac{e^{Nf_0}}{\sqrt{2\pi \kappa}}\bigg( 1 + 
\frac{\alpha_1}{N} 
+ \frac{\alpha_2}{N^2} + \cdots\bigg),
}
\end{equation}
where the $\alpha_i$ are independent of $N$. The first 
two of these coefficients are given by 
\begin{align}\label{alphas}
\alpha_1&=N\!\left(\frac{3B}{\kappa^2}-\frac{15A^2}{2\kappa^3}\right)
\\
\alpha_2&=N^2\left(\frac{15D}{\kappa^3}+\frac{105}{\kappa^4}\bigg(\frac{B^2}{2}-AC\bigg)+\frac{945}{\kappa^5}\frac{A^2B}{2}\right).
\notag
\end{align}
Higher order terms become rapidly more complicated, but 
can be conveniently calculated using symbolic 
algebra software. 
Equation \eqref{final_expansion_ZN} enables approximate 
calculation of the canonical partition function for a finite 
system of $N$ particles, provided that we have some expression 
(either exact or approximate) for the grand-canonical partition function. 

Since equation \eqref{final_expansion_ZN} is an asymptotic expansion we anticipate that adding terms will first lead to a rapid convergence of the partial sum towards the exact value 
of $Z_N$. 
However, beyond a certain point, increasing the number of terms will degrade the numerical accuracy of the partial sum and lead eventually to a divergence. 
Optimal results are to be obtained by truncating the series 
at its `least term' \cite{dingle} and it may be hoped, but 
is by no means certain, that the optimal truncated series can 
be obtained before calculation of the coefficients $\alpha_i$ becomes impractical.
A further complication is that the position of the least term within the series will not necessarily be the same for different values of $N$.

The asymptotic expansion of the canonical free energy follows by taking the logarithm 
of equation \eqref{final_expansion_ZN} and ordering terms 
according to their scaling with $N$. 
We thus obtain 
\begin{align}\label{free_energy_N}
F_N &= \Omega - \mu N  
+ \frac{1}{2}k_BT\ln\left(
2\pi \frac{\partial 
\langle N\rangle}{\partial \beta\mu}
\right)
\notag\\
&+ k_BT \left( 
\frac{\alpha_1}{N} + \frac{(\alpha_2-\frac{1}{2}\alpha_1^2)}{N^2}
\,+\, \cdots
\right).
\end{align}
In the thermodynamic limit the third term scales as 
$\ln(N)$ and so remains finite, but negligable 
in magnitude when compared to the first two terms, 
which are both extensive quantities. 
All higher-order terms vanish as 
$N\!\rightarrow\!\infty$. 
Equation \eqref{free_energy_N} shows explicitly the emergence 
of the Legendre transform as a lowest-order approximation.

We note that the first three terms in equation \eqref{free_energy_N} 
can already be found in a few scattered locations throughout the statistical mechanics literature (see e.g.~\cite{Lebowitz}).  
The reason for this appears to be that these leading terms can be fortuitously obtained using two distinct approximations, both of which are convienient to implement but hard to justify. 
The first, and most dubious, of these consists of replacing the sum 
over particle number in equation \eqref{forwards} by an integral, thus blindly assuming that the sharp particle number 
$N$ can be treated as a continuous variable.  
This approach reproduces the zero-order term in equation \eqref{final_expansion_ZN}, namely the prefactor 
$e^{N f_0}/(2\pi a)^{\frac{1}{2}}$, and thus correctly 
captures the first three terms in equation 
\eqref{free_energy_N}, but generates erroneous higher-order terms. 
Further details of the `continuous $N$' approximation can be found in appendix \ref{continuous}.
The second, perhaps more reasonable, approach is to assume 
a linear path for evaluating the back transform \eqref{backwards}.  
The integration is then performed  
in the imaginary direction along the path 
$\lambda\!=\!x_0 + iy$, implicitly assuming that the contribution 
from closing the path around some large loop around the origin does not contribute to the integral.
As for the continuous $N$ approximation, 
higher order terms in \eqref{free_energy_N} are given incorrectly. 
This indicates that at order $N^{-1}$ and beyond 
in \eqref{free_energy_N} (equivalently, terms involving the $\alpha_i$ 
in equation \eqref{final_expansion_ZN}) are sensitive 
to the curvature of the contour $\mathcal{C}^*$. 
The circular contour identified in subsection 
\ref{steepest} correctly captures this curvature effect 
for the lowest order terms.

\subsection{Yang-Lee zeros}\label{YangLee}

Since the canonical partition functions are positive real numbers, the polynomial expression \eqref{forwards} 
for $\Xi$  will have $N_{\text{max}}$ zeros which, 
by the `complex conjugate root theorem' will appear
in complex-conjugate pairs, removed from the real, 
positive axis (if $N_{\text{max}}$ is odd, then at 
least one of the roots will lie on the negative real axis).
These are the famous Yang-Lee zeros of statistical 
mechanics, which encode important information about the phase behavior of the system \cite{YangLee1,YangLee2}. 
As the system size increases, the Yang-Lee zeros 
become more numerous and accumulate along well-defined 
curves in the complex plane.
Using very general arguments Yang and Lee showed that if a zero approaches the real 
axis at fugacity $x_c$, then the system will undergo a 
phase transition at precisely this fugacity in the thermodynamic limit. 
The distribution of zeros in the complex plane and the 
way in which they approach the positive real axis on 
increasing $N$ provides information about finite-size 
scaling and critical exponents \cite{Bena,Flindt}. 
In recent work Peng {\it et al.}  have shown 
that it is even possible to (indirectly) measure the location 
of the Yang-Lee zeros in experiment \cite{Peng}.

The success of the steepest descents method relies on the 
fundamental assumption that when integrating along the 
optimal contour $\mathcal{C}^*$ the dominant contribution to the integral comes from the region around the saddle 
point. 
It is then assumed that a Taylor 
expansion of $f$ about the saddle will yield a reasonable 
approximation to the integrand along the entire contour. 
However, since the function $f$ depends on the logarithm 
of $\Xi$, see equation \eqref{f_function}, it will clearly
exhibit a divergence at each of the Yang-Lee zeros. 
These singular points impact the asymptotic evaluation of 
\eqref{backwards} in two distinct ways. 
Firstly, it is well-known that there is a close  
connection between the Yang-Lee zeros and fluctuations in particle number (c.f.~the connection mentioned above between 
the zeros and critical exponents).
The factorization approximation leading from equation \eqref{new_taylor} to equation \eqref{factor_series} relies 
on neglecting fluctuations,  
$\langle N^n\rangle\!-\!\langle N\rangle^n\!\approx\!0$, and 
can thus be expected to break down in the vicinity of 
a Yang-Lee singularity.
This has the consequence that the true $\mathcal{C}^*$ will deviate from the approximate circular form when 
approaching a singularity.
Secondly, in the vicinity of a Yang-Lee zero a truncated 
Taylor expansion of the diverging function $f$  
will clearly be inadequate. 
The extent to which this problem is mitigated by deviations 
of the true $\mathcal{C}^*$ away from Yang-Lee singularities 
remains an open question. We will return to this point 
when considering our numerical test-case in subsection \ref{yanglee}.

\subsection{Asymptotic expansion for the one-body density}

We next consider the asymptotic expansion of the 
one-body density. 
In this case the forwards transformation 
is given by an expression very similar to \eqref{forwards}, 
namely 
\begin{equation}\label{forwards_density}
\Xi(\lambda)\rho({\bf r}\,;\lambda) 
=
\sum_{N=0}^{\infty} \lambda^N Z_N \rho_{N}({\bf r}),
\end{equation}
where $\rho$ is the grand-canonical one-body 
density calculated at fugacity $\lambda$ and 
$\rho_{N}$ is the canonical one-body density for a system of strictly $N$ particles.

Arguments analogous to those given in appendix \ref{proof_integral} generate the following back-transform
\begin{equation}\label{backwards_density}
Z_N \rho_{N}({\bf r}) = \frac{1}{2\pi i}
\oint\limits_{\mathcal{C}^*} d\lambda\, \frac{\Xi(\lambda)\,\rho({\bf r}
\,;\lambda)}{\lambda^{N+1}}.
\end{equation}
The procedure to deal with this integral closely follows that 
already presented in subsection \ref{asymptotic_partition}. 
We first rewrite equation \eqref{backwards_density} in the 
following form
\begin{equation}\label{laplace_form_rho}
Z_N \rho_{N}({\bf r}) = \frac{1}{2\pi i}
\oint\limits_{C} d\lambda\, 
\rho({\bf r}
\,;\lambda)\frac{e^{N f(\lambda)}}{\lambda}
\,.
\end{equation}
Taylor expansion of $f$ around the saddle point 
$x_0$ then yields
\begin{align}\label{partial_expansion_density}
\rho_{N}({\bf r}) Z_N
=&
\frac{e^{Nf_0}}{2\pi}\int_{-\pi}^\pi d\phi \,
\rho({\bf r};\lambda)
e^{-\frac{1}{2}\kappa\phi^2 }
\;\;\times\notag\\
&\times\bigg(e^{iA\phi^3+B\phi^4+iC\phi^5+D\phi^6\cdots}
\bigg),
\end{align}
where we have again used the circle approximation to the contour $\mathcal{C}^*$. 
We next perform a Taylor expansion of the grand-canonical density about the saddle point 
\begin{align}\label{taylor_grand_partition_function}
&\rho(\textbf{r};\lambda)
=
\rho(\textbf{r};x_0)
\;+\;
\left(\lambda-x_0\right)\frac{\partial\rho(\textbf{r};\lambda)}{\partial\lambda}\bigg|_{x_0}
\\
&+\;
\frac{\left(\lambda-x_0\right)^2}{2}\frac{\partial^2\rho(\textbf{r};\lambda)}{\partial\lambda^2}\bigg|_{x_0}
\!\!+\;
\frac{\left(\lambda-x_0\right)^3}{6}\frac{\partial^3\rho(\textbf{r};\lambda)}{\partial\lambda^3}\bigg|_{x_0}+\cdots.
\notag
\end{align}
Insertion of the Taylor expansion 
\begin{align}
(\lambda-x_0) &= x_0(e^{i\phi}-1)
\notag\\
&\approx x_0(i\phi+\frac{1}{2}(i\phi)^2+\frac{1}{6}(i\phi)^3+\dots),
\end{align} 
into equation \eqref{taylor_grand_partition_function}
and then collecting powers of $\phi$ yields the phase angle expansion of the grand-canonical one-body density 
\begin{align}\label{grand_density_expansion}
    \rho({\bf r}\,;\lambda)
    =
    \rho({\bf r};x_0) +ia({\bf r})\phi + b({\bf r})\phi^2 + ic({\bf r})\phi^3
\notag\\    
    + d({\bf r})\phi^4 
    + ig({\bf r})\phi^5 
    + h({\bf r})\phi^6 
    + \dots
    \,,
\end{align}
The position-dependent coefficients, $a,b\ldots$ ,  
are functions of derivatives of the grand-canonical density. 
The first few are given by 
\begin{align}\label{asymptotic_coefficients_lowercase}
a(\textbf{r})&=P^{(1)}(\textbf{r}),
\\
b(\textbf{r})&=-\frac{1}{2}\left(P^{(1)}(\textbf{r}) + P^{(2)}(\textbf{r})\right),
\notag\\
c(\textbf{r})&=-\frac{1}{6}\left(P^{(1)}(\textbf{r}) + 
3P^{(2)}(\textbf{r}) + P^{(3)}(\textbf{r})\right),
\notag\\
d(\textbf{r})&=\frac{1}{24}\left(P^{(1)}(\textbf{r}) + 
7P^{(2)}(\textbf{r}) + 
6P^{(3)}(\textbf{r}) + 
P^{(4)}(\textbf{r})\right),
\notag
\end{align}
where we have defined 
\begin{equation}
P^{(n)}(\textbf{r})=\left( x\frac{\partial}{\partial x}\right)^{\!\!n}\!\rho(\textbf{r})\bigg|_{x=x_0}
\end{equation}
Taylor expansion of the exponential containing the upper-case coefficients $A, B, C...$ in equation \eqref{partial_expansion_density}, using 
equation \eqref{grand_density_expansion} 
and extending the range of integration yields a sum of Gaussian integrals. 
Regrouping the resulting terms according to their 
dependence on $N$ we obtain our final form for the asymptotic expansion 
\begin{equation}\label{final_expansion_density}
\boxed{
    \rho_N({\bf r}) =\rho({\bf r}\,;\lambda)\,+
\frac{1}{Z_N}    
\frac{e^{Nf_0}}    
{\sqrt{2\pi\kappa}}\left(
\frac{\Gamma_1(\textbf{r})}{N} + 
\frac{\Gamma_2(\textbf{r})}{N^2} + 
\dots\right), 
}
\end{equation}
where $Z_N$ can be approximated by 
the asymptotic series \eqref{final_expansion_ZN}.
The first $\Gamma$ coefficients are given by 
\begin{align}\label{asymptotic_coefficients_gamma}
\Gamma_1&\!=\!N\left(\frac{b}{\kappa}-\frac{3aA}{\kappa^2}\right),
\\
\Gamma_2&\!=\!N^2\!\left(
\frac{3d}{\kappa^2}+\frac{15(bB\!-\!cA\!-\!aC)}{\kappa^3}-\frac{105(bA^2\!+\!2aAB)}{2\kappa^4}\right),
\notag
\end{align}
where spatial arguments have been suppressed for conciseness. 
The calculation of higher-order coefficients is a straightforward but tedious exercise. 
Although not obvious from casual inspection of the 
expressions \eqref{asymptotic_coefficients_gamma}, 
the coefficients $\Gamma_i$ all have spatial integral 
equal to zero. 
This property ensures the correct normalization of the 
asymptotic approximation \eqref{final_expansion_density} 
at any order of truncation.

As a final comment, we mention that the Yang-Lee zeros can be expected to play an 
equally important role for the asymptotic expansion of 
the one-body density as for the partition function, since both arise from essentially the 
same scheme.

\subsection{Relation to the method of Gonz\'alez}\label{bob_method}

In the late 1990's Gonz\'alez {\it et al.} derived the following expansion 
for the canonical density profile 
\cite{Evans_canonical_letter,Evans_canonical_JCP}
\begin{equation}\label{eq_evans_expansion}
    \rho_{N}({\bf r})
    = 
    \rho({\bf r}) 
    + g_1({\bf r}) 
    + g_2({\bf r})+\dots,
\end{equation}
where the position-dependent terms $g_i$ depend solely on 
grand-canonical input and scale with particle number 
$\sim\!N^{-i}$. 
The first two correction terms in equation \eqref{eq_evans_expansion} 
were found to be given by
\begin{align}\label{EvansCoefficient1}
g_1({\bf r})=-\frac{\langle\, 
(N - 
\langle N\rangle
)^2\rangle}{2}
\frac{\partial^2\rho({\bf r})}{\partial
\langle N\rangle^{\,2}},
\end{align}
and 
\begin{align}\label{EvansCoefficient2}
\!\!\!g_2({\bf r}) =-\frac{\langle 
(N\!-\!\langle N\rangle )^2
\rangle}{2}\frac{\partial^2g_1({\bf r})}{\partial
\langle N\rangle^{\,2}}
-\frac{\langle (N\!-\!
\langle N\rangle )^3\rangle}{6}\frac{\partial^3\rho({\bf r})}{\partial
\langle N\rangle^{\,3}}, 
\end{align}
respectively (see equations $2.10$ and $2.11$ in Reference \cite{Evans_canonical_JCP}).
All grand-canonical quantities 
appearing on the right hand-side of 
\eqref{EvansCoefficient1} and 
\eqref{EvansCoefficient2} are evaluated at the fugacity for which $\langle N\rangle\!=\!N$ (identified as the saddle-point condition in our asymptotic approach). 
When truncated at order $N^{-2}$ the expression \eqref{eq_evans_expansion} was shown to predict 
profiles for hard-spheres confined to a spherical cavity 
in good agreement with canonical Monte-Carlo simulation data.
In appendix \ref{evans_expansion} we discuss the 
recursive back-substitution method used in References 
\cite{Evans_canonical_letter} and \cite{Evans_canonical_JCP}.

Following some algebra, the derivatives with respect to the 
average particle number and the fluctuation coefficients 
in equations \eqref{EvansCoefficient1} and  
\eqref{EvansCoefficient2} can be rewritten in terms of derivatives with respect to the fugacity. 
Further manipulation of the resulting expressions then reveals 
that equations \eqref{eq_evans_expansion}, 
\eqref{EvansCoefficient1} and \eqref{EvansCoefficient2} 
derived by Gonzal\'ez {\it et al.} are, in fact, completely equivalent to the leading-order terms in the asymptotic expression \eqref{final_expansion_density} derived in the previous subsection. 
The recursive back-substitution method 
of Gonz\'alez {\it et al.} thus reproduces the correct 
asymptotic expansion of the canonical one-body density 
to order $N^{-2}$.
It seems to us a nontrivial observation that 
these earlier findings, calculated using a somewhat 
obscure recursive method, can only be reproduced when evaluating the back transform \eqref{backwards} along a sufficiently well chosen contour.
Neither the {\it ad hoc} `continuous $N$' approximation 
nor the linear contour mentioned earlier can achieve this.    
 
For the purpose of illustration we will briefly demonstrate 
the equivalence of 
the leading-order correction term, $g_1$, to the term involving $\Gamma_1$ in equation \eqref{final_expansion_density}. 
An analogous calculation, which we omit here as it is more 
lengthy but not more instructive, shows the equivalence of the second-order terms from the two approaches.
We note that the work of Gonz\'alez {\it et al.} 
makes no reference 
to a complex fugacity and thus all calculations presented for the remainder of this subsection are performed 
on the real axis, $\lambda\!=\!x$.

\begin{figure*}[t!]
\begin{minipage}[b]{0.45\linewidth}
\hspace*{0cm}\includegraphics[width=\textwidth]{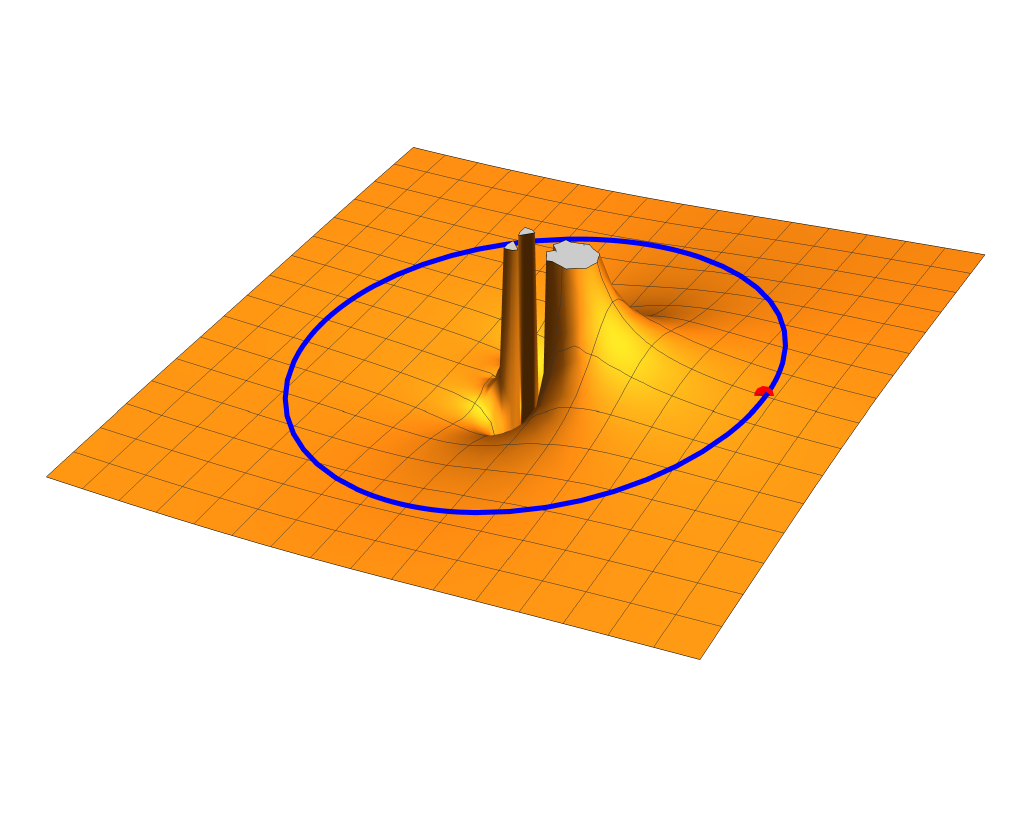}
\end{minipage}
\hspace{0.2cm}
\begin{minipage}[b]{0.45\linewidth}
\centering
\includegraphics[width=0.9\textwidth]{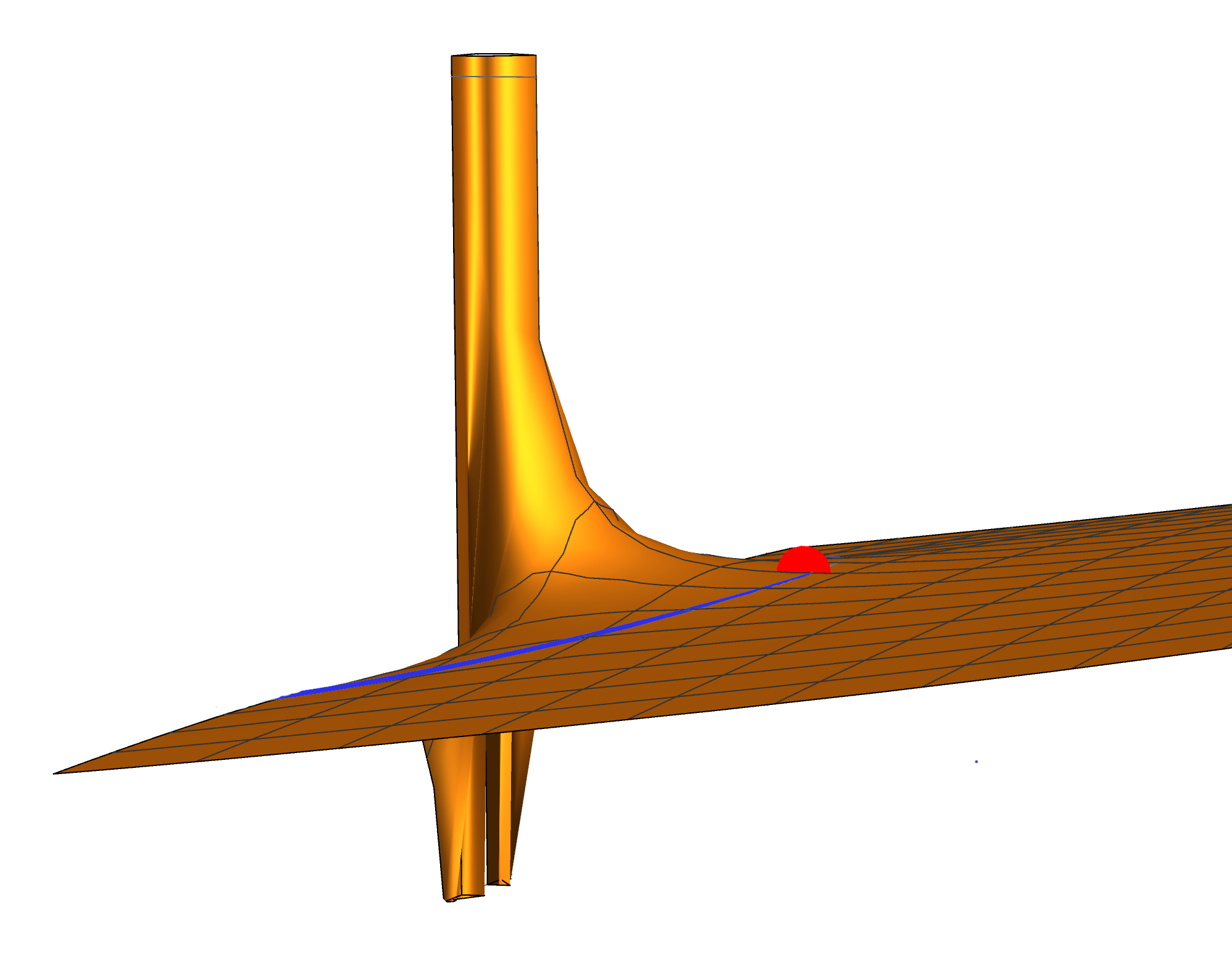}
\end{minipage}
\caption{Visualizing the integrand of the back-transform, equation \eqref{backwards}.
{\bf Left panel:} The quantity $\mathfrak{Re}$ $(\Xi/\lambda^{N+1})$ for $N=3$ in the complex plane, highlighting the saddle point (Red point) and the circular contour we 
use in our calculations (blue curve)
{\bf Right panel:} The same surface viewed from a different angle to better view the saddle point. 
\label{surface}
}
\end{figure*}

We begin by rewriting $g_1$ using only derivatives with 
respect to $x$. This exercise in partial 
differentiation yields
\begin{equation}\label{g1_equation}
    g_1({\bf r})=-\frac{x}{2}
    \left(\!\frac{\partial\langle N\rangle}{\partial x}
    \!\right)^{\!\!\!-2}\!\left(\frac{d\langle N\rangle}{dx}\frac{d^2\rho({\bf r})}{dx^2}-\frac{d^2\langle N\rangle}{dx^2}
    \frac{d\rho({\bf r})}{dx}\right).
\end{equation}
We next express the derivatives of $\langle N\rangle$ appearing in equation \eqref{g1_equation} in terms of derivatives of the function $f$. 
The required identities are
\begin{align}
\frac{\partial \langle N\rangle }{\partial x}
\,\bigg|_{x_0}
\!\!\!&=
Nx_0f^{(2)}(x_0),
\\
\frac{\partial^2 \langle N\rangle }{\partial x^2}
\,\bigg|_{x_0}
\!\!\!&=
N\big(x_0f^{(3)}(x_0)+2f^{(2)}(x_0)\big),
\end{align}
which follow from equations \eqref{Nbar_definition} and \eqref{f_function}.
The superscript on $f$ indicates the 
$n^{\text{th}}$ order derivative with respect to $x$. 
The correction term $g_1$ thus takes the form
\begin{equation}\label{to_prove}
\begin{aligned}
    g_1({\bf r})=&-\frac{1}{2}\frac{1}{Nf^{(2)}(x_0)}\!\left(\frac{\partial^2\rho({\bf r})}{\partial x^2}\right)_{\!x_0}
    \\
    &+\frac{1}{2}\frac{x_0f^{(3)}(x_0)+2f^{(2)}(x_0)}{Nx_0(f^{(2)}(x_0))^2}
    \!\left(\frac{\partial\rho({\bf r})}{\partial x}\right)_{\!x_0}.\\
\end{aligned}
\end{equation}
On the other hand, the leading-order correction term in our asymptotic expansion \eqref{final_expansion_density} is given by
\begin{equation}\label{first_term}
\frac{1}{Z_N}\frac{e^{Nf_0}}{\sqrt{2\pi\kappa}}\frac{\Gamma_1}{N}
\;=\;
\frac{1}{Z_N}
\frac{e^{Nf_0}}{\sqrt{2\pi\kappa}}\left(\frac{b}{\kappa}-\frac{3aA}{\kappa^2}\right).
\end{equation}
From equation \eqref{final_expansion_ZN} we have, to leading order, that the canonical partition function is given by
\begin{equation}
    Z_N=\frac{e^{Nf_0}}{\sqrt{2\pi\kappa}}.
\end{equation}
Direct substitution of the above expression for $Z_N$ as well as the expressions for $a, b$ and $A$ into 
\eqref{first_term} exactly reproduces equation 
\eqref{to_prove}. 
We have thus shown that the $g_1$ term in the expansion 
of Gonz\'alez {\it et al.} is equivalent to the 
leading-order correction term of our asymptotic expansion. 

\section{Application to 1D hard-rods}\label{section_rods}

We will next investigate the asymptotic expressions obtained above using an exactly soluble model, namely hard-rods in one spatial 
dimension, confined between two hard-walls. For this system 
the canonical partition functions and thus (via equation \eqref{forwards}) the grand-canonical partition function are given by simple analytic expressions. 
For the one-body density, exact grand-canonical profiles can 
be calculated by variational minimization of the Percus 
hard-rod free-energy functional \cite{Evans92,PercusOriginal}. 
Application of exact matrix inversion then yields the canonical profile for all values of $N$ 
\cite{ExactCanonical}. 
The one-dimensional confined rod model is thus a rare case 
for which we have a complete 
and exact description of the thermodynamics and one-body density in both the canonical and grand-canonical ensembles.

The canonical partition function of $N$ rods confined between 
two hard walls separated by a distance $L$ is given exactly by
\begin{equation}\label{exact_rod_ZN}
Z_N = \frac{1}{N!}
\left(L-N\right)^N, 
\end{equation}
where we have set both the thermal wavelength and the rod length equal to unity. 
The corresponding grand-canonical partition function is 
then given by the definition \eqref{forwards} and can be 
directly evaluated for any complex value of the fugacity.

Grand-canonical density profiles can be determined 
by solving the Euler-Lagrange equation 
\eqref{EL} using the following exact expression for  the 
excess Helmholtz free energy 
\begin{equation}\label{percus_functional}
    F^{ex}[\rho]=-k_BT\int_{-\infty}^{\infty} n_0(x)\ln (1-n_1(x))dx
\end{equation}
where the weighted densities $n_0$ and $n_1$ are generated 
from the convolution integrals
\begin{equation}
    \begin{aligned}
    &n_{\alpha}(x) = \int_{-\infty}^{\infty}\rho(x')
    w_{\alpha}(|x-x'|)dx',
    \end{aligned}
\end{equation}
with geometrically-based weight functions given by
\begin{align}
    w_0(x) &= \frac{1}{2}\Big( 
    \delta(x-R)+\delta(x+R)\Big)
\\
w_1(x)&=\begin{cases}
    1, & \text{if $-R < x < R$}\\
    0, & \text{otherwise}\,,
  \end{cases}
\end{align}
where $R\!=\!1/2$ in our chosen units of length. 
Calculating grand-canonical profiles at $N_{\text{max}}$ different values of the fugacity enables the definition \eqref{forwards_density} to be cast as a matrix multiplication. Numerical inversion then yields all 
canonical density profiles, $\rho_N$, for 
$N\!=\!1,\ldots ,N_{\text{max}}$, as described in 
Reference \cite{ExactCanonical}.

\begin{table}[t!]
%
    \centering
    \begin{tabular}{|c||c|c||c|c||c|c|}
    \hline
        $N$ & $x_0$ & $Z_N^{exact}$ & $Z_N^{(1)}$  & $\varepsilon_1$ & $Z_N^{(2)}$ & $\varepsilon_2$\\
        \hline
         1& 0.240534 & 5.9 & 6.39083 & 7.68\% & 6.20258 & 4.87\%\\
         2& 0.760902 & 12.005 & 12.4889 & 3.87\% & 11.4915 & 4.27\%\\
         3& 2.16724 & 9.8865 & 10.1417 & 2.51\% & 9.46997 &4.21\% \\
         4& 7.8077 & 2.9470 & 3.00396 & 1.89\% & 3.0228 & 2.50\%\\
         5& 66.3881 & 0.2063 & 0.206539 & 0.11\% & 0.309272 &  33.2\%\\
         \hline
    \end{tabular}
\caption{Data for a system of six hard-rods confined between hard walls with separation $L\!=\!6.9$. 
We give the saddle point fugacity, 
$x_0$, the exact partition function, $Z_N$, 
and both the first- and second-order asymptotic approximation, 
$Z_N^{(1)}$ and $Z_N^{(2)}$, respectively. 
The percentage error of these approximations is 
given by $\varepsilon_{1,2}$.  
Data for the maximally packed state $N\!=\!6$ are omited, since the results are not reliable. 
\label{table1}}    
\end{table}

Let us now choose a specific and representative example by fixing $L=6.9$. 
Due to this confinement the system can admit at most six rods and 
the grand partition function \eqref{forwards} is a 
finite polynomial. 
For each value of $N$ we can determine exactly the corresponding canonical partition function $Z_N$ using 
\eqref{exact_rod_ZN} as well as the position of the saddle point 
$x_0$ using the saddle condition 
\eqref{saddle_condition}, see the values given in Table \ref{table1}. 
We note that the fugacity of the saddle point corresponding to 
the most densely packed state, namely 
$\langle N\rangle\!=\!6$, is so large that it is practically 
unattainable by numerical minimization of the grand 
potential. 

\subsection{Numerical results: Partition function}

We begin with a quick visual inspection the integrand 
of equation \eqref{backwards}, namely the function 
$\Xi/\lambda^{N+1}$. Despite the many applications of 
saddle-point approximation in the physical and mathematical literature, 
it is exceedingly rare to find actual visualizations 
of the saddle for any concrete problem. 
We thus show in 
Fig.~\ref{surface} a surface plot of the real part of the 
integrand in the complex plane for the case
$N\!=\!3$. 
As anticipated, we indeed observe a saddle point on the real 
axis, which is located at $x_0\!=\!2.167$ and indicated in 
the Figure by a red dot. 
The blue curve shows the approximate circular contour we 
employ in our calculations.

Table \ref{table1} gives the asymptotic approximation to 
the canonical partition functions, obtained from equation \eqref{final_expansion_ZN}, together with their exact values 
and the location of the saddle points.
For $N\!\le\!5$ truncation at first-order provides a 
good approximation to the exact value of $Z_N$, with  
an error which decreases with increasing 
$N$. 
This is the behavior one would expect from 
a valid asymptotic expansion. 
The results to second-order are, however, less 
satisfactory and, with the exception of the case 
$N\!=\!1$, degrade the accuracy of the approximation.  
We suspect that already at second-order the output of the theory becomes sensitive to the choice of contour and that 
the approximate circular contour is already insufficient to 
generate the true second order terms which would result from 
employing the optimal contour $\mathcal{C}^*$. 
In the following subsection we will analyse these contour 
differences in more detail, but they are closely connected with the presence of Yang-Lee zeros.

\begin{table}[b!]
\begin{tabular}{|c|c|c|c|c|c|}

    \hline
        $x^{\text{YL}}_1$ & $x^{\text{YL}}_2$ & $x^{\text{YL}}_3$ & $x^{\text{YL}}_4$ & $x^{\text{YL}}_5$ & $x^{\text{YL}}_6$\\
        \hline
        -0.404 & -0.545 & -0.940 & -2.304 & -10.728 & -264.172\\
        \hline
    \end{tabular}
\caption{For a system of six hard-rods confined between two
hard walls with separation $L\!=\!6.9$ all six Yang-Lee zeros lie
on the negative real axis in the complex fugacity plane. 
\label{table2}}    
\end{table}

A well-known property of asymptotic series is that increasing the number of terms generates partial sums which initially converge towards the exact result but, beyond a certain 
point, begin to diverge 
as further terms are added \cite{dingle}. 
Truncation at the so-called `least term' provides an optimal approximation 
(although further improvements can still be made 
\cite{hyperasymptotics}).
This suggests an alternative interpretation of the 
data in Table \ref{table1}, namely that the asymptotic 
series has reached its least term already at first-order 
and that this then represents the optimal truncation.

\begin{figure*}
\centering  
\includegraphics[width=0.98\textwidth]{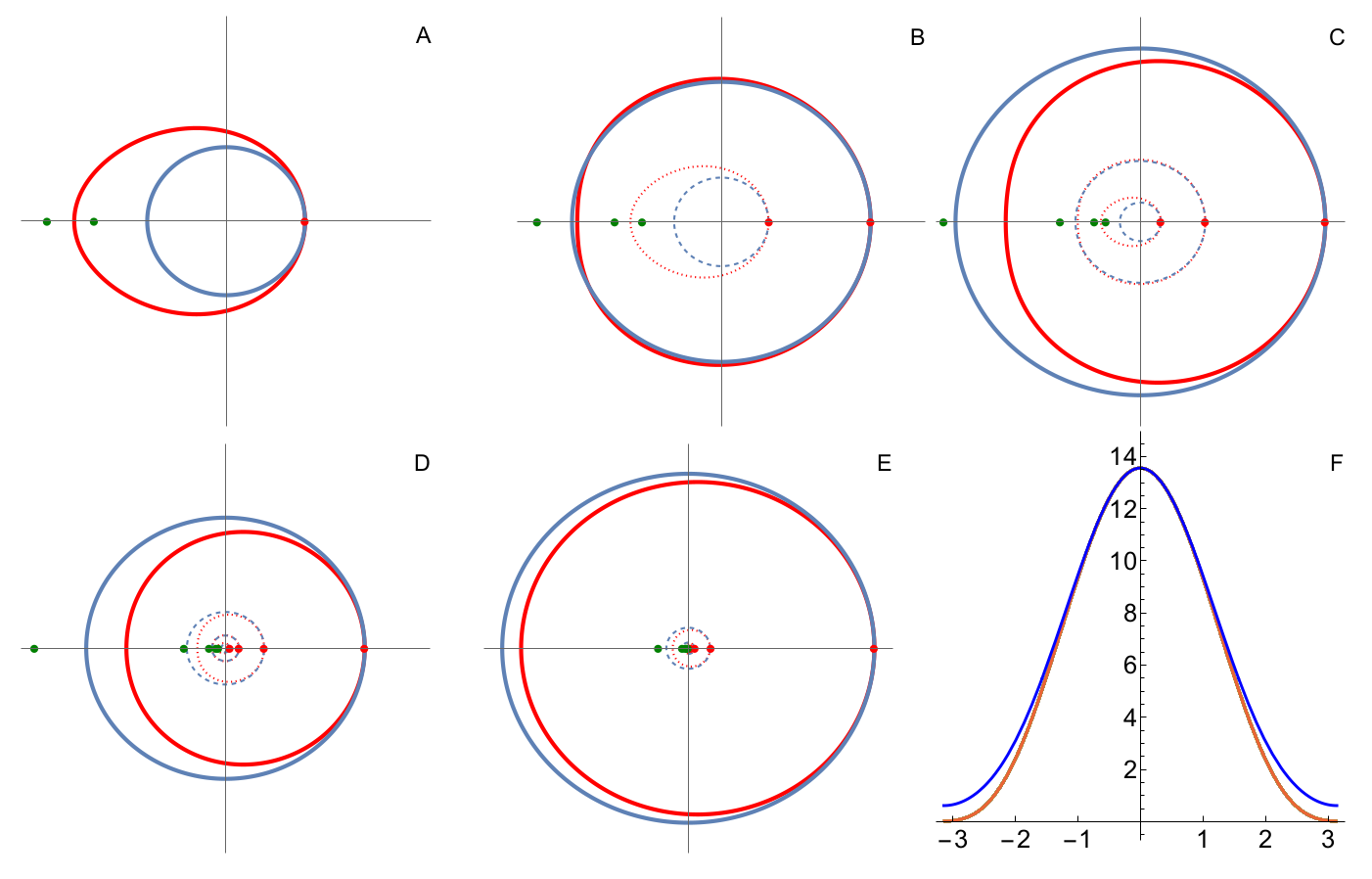}
\caption{Figures A to E show the true optimal contour 
$\mathcal{C}^*$ (red) and the approximate circular contour (blue) for 
(A) $N\!=\!1$, 
(B) $N\!=\!2$, 
(C) $N\!=\!3$, 
(D) $N\!=\!4$, 
(E) $N\!=\!5$, in the complex fugacity plane. Red dots represent the saddle points, and green dots represent Yang-Lee zeroes. 
The dotted and dashed curves in each of the Panels 
(B)-(E) simply show the contours from the preceding Panels 
for visual reference as we zoom out.
For small phase-angles the two contours show good agreement but then deviate as the Yang-Lee zeros are approached. 
Note how the true optimal contour deforms to enclose 
exactly $N$ Yang-Lee zeros.
Figure (F) shows $\exp(N f_{\text{re}})$ for the case $N\!=\!1$ 
along both the true contour (red) and the circular contour (blue). 
\label{YL_figure}
}
\end{figure*}

\subsection{Yang-Lee zeros and the optimal contour}\label{yanglee}

For the hard-rod system under consideration we find 
that all six Yang-Lee zeros are located on the negative 
real axis in the complex $\lambda$-plane (which is consistent with earlier studies \cite{niemeyer69,niemeyer70}). 
Their locations are listed in Table~\ref{table2}, where 
we observe that the distance of the zeros from the origin 
increases very rapidly as the maximally packed state 
($N\!=\!6$) is approached. 
In Figure~\ref{YL_figure} we plot both the Yang-Lee zeros 
and the saddle points, together with the approximate circular contour (blue curve) and the true optimal contour (red curve). 
The latter was obtained by starting from the 
saddle point and then numerically solving the equation 
$f_{\text{im}}\!=\!0$ to track a path through the  
complex $\lambda$-plane.

For small values of the phase angle, $\phi$, the true contour is well 
approximated by the circle, as might have been anticipated by 
comparing the exact equation \eqref{new_taylor} with the 
approximate form \eqref{factor_series}. 
As the value of $\phi$ is increased the fluctuation terms, 
$\langle N^i\rangle\!-\!\langle N\rangle^i$, neglected in 
equation \eqref{factor_series} begin to play an 
increasingly important role and the true contour deviates 
from the circular form. 
We observe that the true contour always seems to deform 
in such a way that it threads neatly between the Yang-Lee 
zeros in an attempt to maintain as much distance as possible from these singular points. Somewhat more mysteriously, the true contour for any given $N$ deforms to 
enclose precisely $N$ Yang-Lee zeros. 
This behaviour, which is most clearly visible for the case 
$N\!=\!1$, suggests a deep connection between the saddle 
points, the Yang-Lee zeros and the topology of the zero-phase 
contour $\mathcal{C}^*$. 

\begin{figure*}[t!]
\centering
\includegraphics[width=0.95\textwidth]{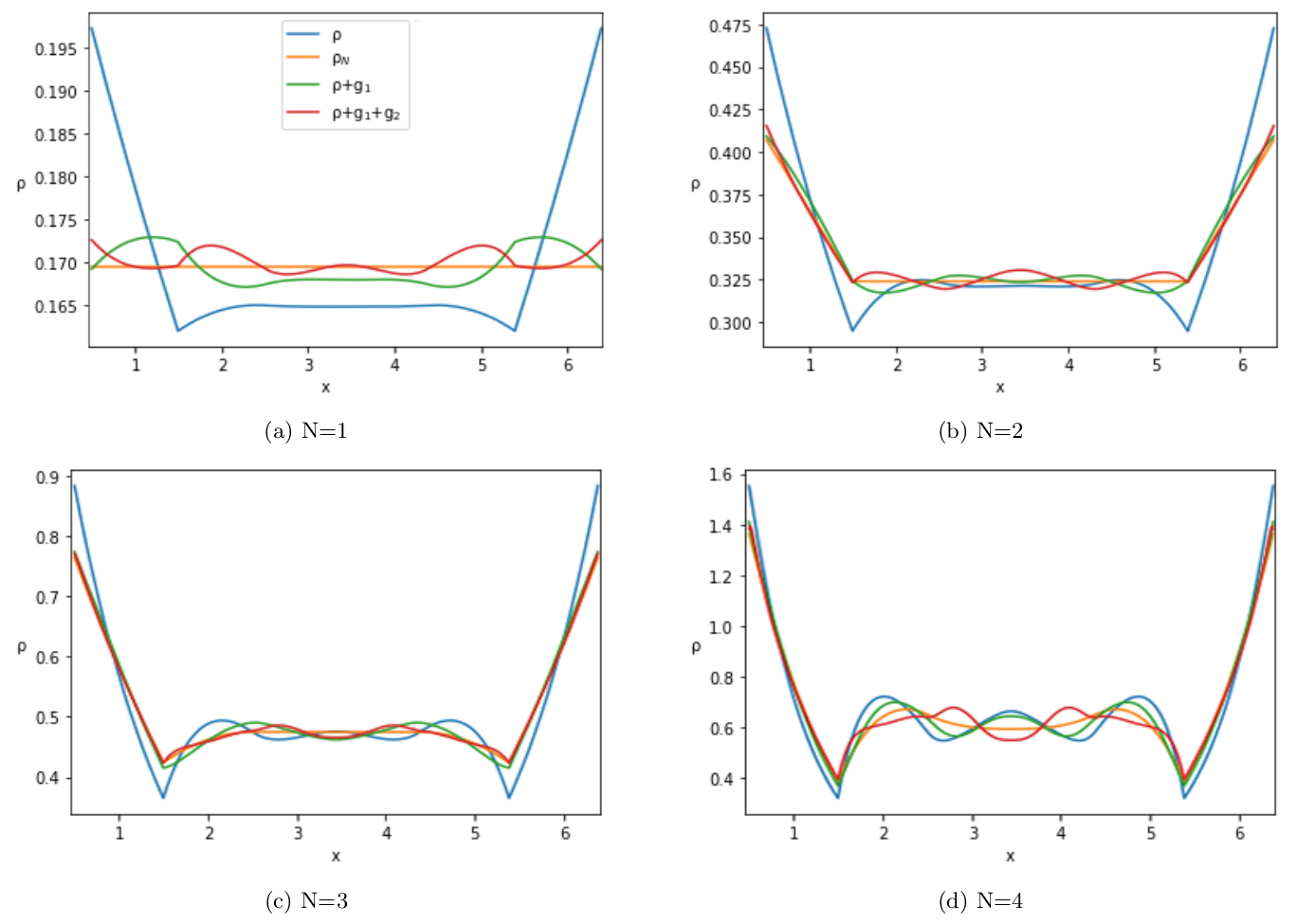}
\caption{
Density profiles for hard-rods confined between two hard walls with separation $L\!=\!6.9$. The maximal number of rods which can fit in the slit is $N_{\text{max}}\!=\!6$. 
Profiles are shown for the 
cases $N\!=\!1,\ldots,4$. For larger values of $N$ the performance degrades and we encounter a high-packing 
breakdown of the asymptotic expansion. The first-order term provides clearly corrects the most serious deficiencies of the grand-canonical profiles, whereas the second order term fails 
to provide a convincing systematic improvement.   
\label{density_figure}}
\end{figure*} 

It seems to us very likely that the deviations of the circular 
contour from the true optimal contour are responsible for 
the disappointing performance of the second-order 
asymptotic trunction detailed in Table \ref{table1}.  
Focusing on the case $N\!=\!1$ we plot in Panel (F) 
the function $e^{N f_{\text{re}}}$, 
which is the key part of the integrand in equation \eqref{factored}. 
Although the circular contour provides a reasonable approximation to $e^{N f_{\text{re}}}$, deviations are clearly visible as the negative real axis is approached. 
This corresponds to the range of phase-angles for which 
the true contour deforms to avoid the
singularity in $f$ arising from the first Yang-Lee zero. 
We recall that the circular contour originates from application of the factorization approximation to the exact series 
\eqref{new_taylor} and thus implicitly assumes that fluctuations are negligable. 
Since the Yang-Lee zeros are associated with increased 
particle-number fluctuations the assumptions underlying 
the circle approximation can be expected to break down 
close to these singular points. 
This suggests that the (nontrivial) relation between the positions of 
the saddle points on the positive real axis 
and the Yang-Lee zeros on the negative real axis can have a 
significant effect on the accuracy of the circle 
approximation. 
Specifically, if $x_0\!\approx\!|x_i^{\text{YL}}|$ for any value of $i$, 
then the approximate circular contour will pass close to the 
zero and will thus be unlikely to provide a good account of the true contour. 

The simplicity of the hard-rod model restricts the 
Yang-Lee zeros to the negative real axis and, although 
their influence is not negligable, they are sufficiently 
distant from the saddle points that the asymptotic method 
can still be applied with reasonable success.  
More realistic model fluids with attractive 
interparticle interactions will certainly present a 
more complicated pattern of zeros. 
For example, simply adding 
a square-well nearest-neighbour attraction to the 
hard-rod model is sufficient to move the zeros off the 
negative real axis \cite{niemeyer69,niemeyer70}.
A clear implication of our findings is that a nontrivial distribution of zeros over the complex fugacity plane 
would  seriously complicate determination of an 
appropriate contour. It therefore seems unlikely that  
the simple circle approximation would be sufficient 
to correctly determine even the leading-order terms 
in the asymptotic expansion for attractive systems.

\subsection{Numerical results: One-body density}

For three-dimensional hard-spheres confined within a spherical cavity  
Gonz\'alez \textit{et al.} obtained accurate profiles 
using the first two 
terms in the expansion \eqref{final_expansion_density} (equation \eqref{eq_evans_expansion} when written in their notation); results which we have indeed verified 
using our own independent codes. 
For some special packing configurations one of the spheres can be quite tightly 
localized at the center of the cavity 
(quasi-0D states) due to interactions 
with its surrounding neighbours. 
These situations were found to have strongly 
ensemble-dependent density profiles and were thus 
used to test the convergence of the approximate canonical expansion.
Although the first-order correction already gave a good account of canonical Monte-Carlo simulation data, it 
remained uncertain that the addition 
of the second-order term could provide a systematic 
improvement over the first-order results \cite{Evans_canonical_JCP}.

The present model system of confined hard-rods presents 
a more demanding test of the asymptotic method than the 
three-dimensional cavity, since ensemble differences are 
more pronounced and the packing constraints on the 
particles are more restrictive. 
In Figure~\ref{density_figure} we investigate the predictions 
of the asymptotic approximation to the one-body density for particle numbers $N=1,\ldots,4$. 
The blue curves are the exact grand-canonical profiles, 
calculated by solving the Euler-Lagrange equation 
\eqref{EL} using the functional \eqref{percus_functional}, whereas 
the orange curves show the exact canonical profiles we 
seek to approximate.  
The green and red curves then give the asymptotic results 
to first- and second-order, respectively,  
corresponding to using the $\Gamma_1$ and $\Gamma_2$ 
terms of equation \eqref{final_expansion_density} 
(equivalently, the $g_1$ and $g_2$ terms in the 
Gonz\'alez expansion \eqref{eq_evans_expansion}). 

We begin by discussing the extreme case of a single rod, 
$N\!=\!1$. In Panel (a) of Figure \ref{density_figure} we observe a large difference between the (trivial) canonical 
density profile and the grand-canonical profile obtained from 
DFT. The latter contains unwanted contributions from 
particle number fluctuations and thus exhibits unphysical 
packing structure. 
Despite the fact that $N\!=\!1$ is certainly not a large value 
of $N$ the asymptotic method performs better than one might expect and shows clear evidence of convergence towards 
the exact canonical result. 
Although the asymptotic approximation introduces unphysical 
oscillations, these both reduce in amplitude and snake more tightly around the exact solution in going from first- to second-order. A similar trend is visible in Panels (b)-(d). 
We note that the `contact regions' of the density profiles, 
$0.5\!<\!x\!<\!1.5$ and $5.4\!<\!x\!<\!6.4$, are very well 
approximated by the asymptotic expansion, whereas the central region exhibits a slower convergence. 
This behaviour is particularly apparent for the case 
$N\!=\!4$, where addition of the second-order correction term only marginally improves on the first-order predication. 
These general trends are fully consistent with the three-dimensional results of Gonz\'alez \textit{et al.} who observed that convergence 
of the density profiles was more rapid close 
to the boundaries than in the center of the cavity. 

For $N\!=\!5$ and $6$ we encounter difficulties to 
obtain reasonable profiles from the asymptotic expansion.
First indications of this breakdown are already apparent 
in the slower convergence of the partial sums for the case 
$N\!=\!4$ (shown in Panel (d) of Figure \ref{density_figure}) 
and become much worse as maximal packing is approached. 
A similar breakdown at large cavity packings was observed 
by Gonz\'alez \textit{et al.} (see page 3684 of Reference 
\cite{Evans_canonical_JCP}). 
It appears that the asymptotic expansion of canonical quantities for confined fluids is subject to two competing 
mechanisms. 
On the one hand increasing the value of $N$ initially leads 
to improved convergence, as would be expected from an 
asymptotic series. 
On the other hand as the value of $N$ approaches 
$N_{\text{max}}$ a second mechanism becomes relevant and 
leads to the high-packing breakdown. 
The consequence is that the expansion exhibits a 
`sweet spot' at intermediate values of $N$, for which our 
expressions perform best. 
The origin of the breakdown and its
relation to the choice of contour remains an open question.

\section{Discussion and conclusions} \label{discussion}

We have developed the method of asymptotics to address 
the problem of calculating the thermodynamics and 
microstructure of confined fluids in the canonical 
ensemble. 
Although by no means the first attempt in this direction 
we think that our formulation clearly exposes the 
fundamental mechanisms determining the success or 
failure of the method, all of which are ultimately related 
to the presence of Yang-Lee zeros. 
Indeed, it is difficult to imagine making the connection 
between asymptotic series and the Yang-Lee zeros from 
either the recursive back-substitution method of 
Gonz\'alez \textit{et al.} 
\cite{Evans_canonical_letter, Evans_canonical_JCP}
or direct matrix inversion \cite{ExactCanonical}, 
since these approaches do not involve the complex 
fugacity plane in which the zeros are to be found. 

Our chosen test-system of one-dimensional hard rods 
provides a good illustration the main 
issues, since for this model the Yang-Lee zeros are sufficiently removed from the saddle points that their influence is not overly dramatic, but still significant 
enough that we can draw some general conclusions. 
In other words, the zeros make their presence felt, but 
not to the extent that they completely invalidate the 
approximations we have employed, as could well be the 
case for systems with more realistic interaction 
potentials. 
Moreover, the fact that the hard rod system is exactly 
soluble removes any confusion which could arise when 
using approximate grand-canonical input; interpretation 
of the asymptotic series is already complicated enough. 
Nevertheless, the interaction of grand-canonical input 
error with the asymptotics of the back-transform 
surely remains both an interesting and neccessary topic for 
future study. 

One-dimensional hard rods do not exhibit a bulk phase transition, which is reflected in the fact that the 
Yang-Lee zeros for this model all lie along the negative 
real axis in the complex fugacity plane. 
For systems with a bulk phase transition this will no longer 
be the case and the presence of nontrivial zeros off the 
real axis will pose serious challenges for the 
asymptotic method.
Even if the location of the zeros could be identified, 
then this would still leave open the dual problem of 
determining the optimal steepest descents contour 
and then actually performing the required integration 
along it. 
Phase transitions are characterized by a thermodynamic limit 
which takes different analytic forms in distinct regions of 
thermodynamic parameter space. 
We thus suspect that the Stokes phenomena will become an important part of the picture; 
the form of the asymptotic expansion changes discontinously
as a thermodynamic parameter (e.g.~the complex fugacity) 
moves across a Stokes (or anti-Stokes) line 
\cite{stokes_original,stokes_berry,stokes_berry2}. 
This behaviour is a consequence of approximating an analytic 
function using simpler multivalued functions. 
It seems likely that the Yang-Lee zeros will turn out to be located 
on some type of Stokes line which separates different forms 
of the asymptotic expansion as the thermodynamic limit is approached.  
Investigations into this potentially rich phenomenology are underway.

Finally, we mention that an analogous asymptotic approach 
can be employed to back-transform canonical observables 
to the microcanonical ensemble. 
In this case the required contour integral is performed 
in the plane of complex inverse-temperature, $\beta$, and  
phase transitions are manifest as zeros 
in the complex $\beta$ plane \cite{FisherZeros}. 
These so-called Fisher zeros approach the positive real axis in the thermodynamic limit in much the same way as the 
Yang-Lee zeros in the present study. 
It is thus clear that the Fisher zeros will 
influence the choice of a suitable integration 
contour and thereby affect the form of any asymptotic expansion.

\acknowledgements

We would like to thank S.~M. Tschopp for insightful comments 
and stimulating discussions. 

\appendix

\section{Contour integral for obtaining the canonical partition functions}\label{proof_integral}

The inverse transform \eqref{backwards} is a special 
case of the well-known Cauchy integral. 
To demonstrate the validity of \eqref{backwards} we substitute the definition of the grand 
partition function \eqref{forwards} into the contour 
integral. 
This yields
\begin{align}\label{contour_proof}
    &2\pi i Z_N=\;\oint_C \frac{d\lambda}{\lambda^{N+1}} \;+\;\oint_C \frac{Z_1}{\lambda^{N}}d\lambda+ \cdots
\\    
    &
\cdots+\;\oint_C \frac{Z_{N-1}}{\lambda^2}d\lambda    
+\oint_C \frac{Z_N}{\lambda}d\lambda    
+\;\oint_C Z_{N+1}d\lambda   
\;+\;\cdots
\notag    
\end{align}
All terms have an integrand of the form $Z_j/\lambda^{\alpha}$, where we recall that the canonical partition functions do 
not depend on $\lambda$. 
The simplest and most convenient choice of contour enclosing 
the origin is a circle of radius $x_0$. 
Using the parameterization $\lambda=x_0 e^{i\phi}$ gives $d\lambda=ix_0 e^{i\phi}d\phi=i\lambda d\phi$. 
Each term in the expansion then involves an integral of the following form
\begin{equation}
\begin{aligned}
\oint_C \frac{d\lambda}{\lambda^{\alpha}}
&=i\!\int_0^{2\pi}\!d\phi \,\lambda^{1-\alpha}\\
&=ix_0^{(1-\alpha)}\!\int_0^{2\pi}\!d\phi 
\,e^{i\phi(1-\alpha)}.
\end{aligned}
\end{equation}
This integral yields zero, unless the exponent in the 
exponential vanishes. We thus conclude that 
\begin{equation}
    \oint_C \frac{Z_j}{\lambda^{\alpha}}d\lambda=\left\{ \begin{array}{l}
        2\pi i \,Z_j,\quad \mbox{if } \alpha=1 \\
        0, \quad\quad\;\;\;\mbox{otherwise}.
    \end{array}\right. 
\end{equation}
Inserting this result into \eqref{contour_proof} produces 
the identity, thus proving the inverse formula 
\eqref{backwards}. 
The integral operator
\begin{equation}
\frac{1}{2\pi i}
\oint\limits_{C} d\lambda\, \frac{1}{\lambda^{N+1}}
\,,
\end{equation}
which depends on the parameter $N$, 
thus acts on $\Xi$ to extract the term $Z_N$ from 
the sum \eqref{forwards}.

\section{Treating N as a continuous variable}\label{continuous}

Occasionally in the statistical mechanics literature 
one finds a vague description of how to invert equation \eqref{forwards} by replacing the discrete 
sum over $N$ with a continuum integral. 
The reason for the enduring appeal of this suspicious scheme 
is that it rather easily generates the first three terms 
in the asymptotic expansion of the Helmholtz free energy (although 
higher-order terms are incorrect).
In the following we give some details of this 
`continuous $N$' approximation.

Using the standard relation $F_N\!=\!-k_BT \ln(Z_N)$, where 
$F_N$ is the true canonical Helmholtz free energy for fixed particle number, 
we can rewrite equation 
\eqref{forwards} in the following form
\begin{equation}
\begin{aligned}
    \Xi=\sum_{N=0}^\infty e^{\beta\mu N-\beta F_N}.
\end{aligned}
\end{equation}
If we now close our eyes and simply replace the sum by 
an integral we obtain
\begin{equation}\label{integralN}
    \Xi=\int_0^\infty \!dN\, e^{
    -\mu\beta\left(-N+\frac{1}{\mu}F_N\right)}, 
\end{equation}
where we have factored-out the chemical potential in the exponent so 
that it can be treated as a large parameter. 
Since non-integer values of $N$ are not physical the status 
of equation \eqref{integralN} is obviously questionable. 

To apply Laplace's method we first rewrite the integral 
\eqref{integralN} in the standard form
\begin{equation}\label{integral_N_Laplace}
    \Xi=\int_0^\infty \!dN\, e^{-\mu g(N)}
\end{equation}
where we have defined
\begin{equation}
g(N)=\beta\left(-N+\frac{F_N}{\mu}\right).
\end{equation}
Setting $dg/dN\!=\!0$ locates the saddle point, which 
corresponds to setting $\langle N\rangle\!=\!N$. 
Taylor expansion of the exponent in 
equation \eqref{integral_N_Laplace} around the saddle 
and truncating at second-order then yields the approximation
\begin{equation}\label{N_partition}
\begin{aligned}
    \Xi&\approx e^{\beta(\mu 
    \langle N\rangle-F_N)}\int_0^\infty \!dN\, e^{
\left(N - \langle N\rangle\right)^2    
    \beta F_N^{(2)}} 
    \\
    &=e^{\beta(\mu 
    \langle N\rangle-F_N)}
    \sqrt{\frac{2\pi}{\beta F_N^{(2)}}}
\end{aligned}
\end{equation}
where $F_N^{(2)}$ is the second derivative of $F_N$ with respect to the `continuous' variable $N$. 
The saddle condition implies the standard canonical 
relation
\begin{equation}
\mu = \frac{\partial F_N}{\partial N}
\bigg\rvert_{N=\langle N\rangle}.
\end{equation}
Substitution into equation \eqref{N_partition} then gives 
the following approximation to the grand partition function
\begin{equation}
    \Xi \approx 
    e^{\beta(\mu 
    \langle N\rangle-F_N)}
    \sqrt{2\pi\frac{\partial \langle N\rangle}{\partial \beta\mu}}.
\end{equation}
Taking the logarithm yields 
\begin{equation}
\begin{aligned}
    \ln(\Xi)& 
\approx    
    \langle N\rangle\mu\beta-\beta F_N + \frac{1}{2}\ln\bigg(2\pi\frac{\partial \langle N\rangle}{\partial \beta\mu}\bigg),
\end{aligned}
\end{equation}
which, after trivial rearrangement, yields the final approximation to the Helmholtz free energy
\begin{equation}
F_N = \Omega - \mu N  
+ \frac{1}{2}k_BT\ln\left(
2\pi \frac{\partial 
\langle N\rangle}{\partial \beta\mu}
\right). 
\end{equation}
We thus recover the first three terms in the asymptotic expansion \eqref{free_energy_N}.

\section{The Gonz\'alez expansion method}\label{evans_expansion}

In the late 1990's Gonz\'alez {\it et al.} developed 
a scheme to approximate the canonical one-body density 
profile using grand-canonical input quantities 
obtained from DFT. 
The following derivation is our interpretation of 
that given in References \cite{Evans_canonical_JCP,Evans_canonical_letter}. 
We hope that this does justice to the original work and 
that it clarifies any possible points of confusion.

The starting point of the approach is to write the 
grand-canonical one-body density as a weighted sum of 
canonical one-body densities
\begin{equation}\label{Gonzalez1}
\rho({\bf r})=\sum_{N=0}^{\infty}\rho_N({\bf r})P(N;\lambda),
\end{equation}
where $P(N)$ is the probability to find $N$ particles 
at (real) fugacity $\lambda$. Equation \eqref{Gonzalez1} is simply a rewriting of equation \eqref{forwards_density} in the main text.
For a given value of $\lambda$ the average particle number 
is 
\begin{equation}\label{average_N}
\langle N\rangle=\sum_{N=0}^{\infty}N P(N;\lambda).
\end{equation}
Before proceeding further we find it useful to identify a 
hypothetical one-body density, 
$\widetilde{\rho}_N$. 
This is a generalized 
canonical one-body density for which $N$ is viewed as a 
continuous variable. For integer values of $N$ it 
coincides with the true canonical density. 
Let us now rewrite equation \eqref{Gonzalez1} using 
this new density
\begin{equation}\label{Gonzalez_tilde}
\rho({\bf r})=\sum_{N=0}^{\infty}\widetilde{\rho}_N({\bf r})P(N;\lambda).
\end{equation}
The introduction of, $\widetilde{\rho}$, represents the most 
uncertain aspect of the procedure presented here, since 
it is is not a well-defined statistical mechanical 
quantity for non-integer values of $N$. 
However, this step seems to be necessary to generate a 
useful recursive scheme.

We next Taylor expand the $\widetilde{\rho}$ about the chosen value of $\langle N\rangle$ in powers of the continuous variable $N$. This yields 
\begin{align}\label{Gonzalez2}
&\rho({\bf r})=\sum_{N=0}^{\infty}P(N;\lambda)
\bigg(
\widetilde{\rho}_{\langle N\rangle}({\bf r}) 
+
\Big(\!N \!-\!\langle N\rangle\Big)
\frac{\partial \widetilde{\rho}_N({\bf r})}{\partial N}\bigg|_{\langle N\rangle}
\notag\\
&\hspace*{1.7cm}+\!
\frac{1}{2}\Big(\!N \!-\!\langle N\rangle\Big)^{\!2}
\frac{\partial^{2}\widetilde{\rho}_N({\bf r})}{\partial N^2}\bigg|_{\langle N\rangle}
\,+\;\cdots\;
\bigg).   
\end{align}
where the partial derivatives on the right-hand side are evaluated at $N\!=\!\langle N\rangle$ and 
$\widetilde{\rho}_{\langle N\rangle}\!\equiv\!
\widetilde{\rho}_{N=\langle N\rangle}$. 
Performing the sum in equation \eqref{Gonzalez2} then gives
\begin{align}\label{Gonzalez3}
\rho({\bf r})\,=\,&
\widetilde{\rho}_{\langle N\rangle}({\bf r}) 
\,+\, 
\frac{1}{2}
\left\langle \left( N \!-\!\langle N\rangle\right)^{2}
\right\rangle
\frac{\partial^{2}\widetilde{\rho}_N({\bf r})}{\partial N^2}\bigg|_{\langle N\rangle}
\notag\\
&+\,
\frac{1}{6}
\left\langle \left( N \!-\!\langle N\rangle\right)^{3}
\right\rangle
\frac{\partial^{3}\widetilde{\rho}_N({\bf r})}{\partial N^3}\bigg|_{\langle N\rangle}
+\;\cdots\;
.  
\end{align}
We seek to find the first term on the right hand side of equation \eqref{Gonzalez3}. Trivial rearrangement yields
\begin{align}\label{Gonzalez4}
\widetilde{\rho}_{\langle N\rangle}({\bf r}) 
\,=\,&
\rho({\bf r})
\,-\, 
\frac{1}{2}
\left\langle \left( N \!-\!\langle N\rangle\right)^{2}
\right\rangle
\frac{\partial^{2}\widetilde{\rho}_N({\bf r})}{\partial N^2}\bigg|_{\langle N\rangle}
\notag\\
&\hspace*{-1cm}-\,
\frac{1}{6}
\left\langle \left( N \!-\!\langle N\rangle\right)^{3}
\right\rangle
\frac{\partial^{3}\widetilde{\rho}_N({\bf r})}{\partial N^3}\bigg|_{\langle N\rangle}
+\;\cdots\;
.  
\end{align}
We next assume that we can make the replacement  
\begin{equation}\label{replacement}
\frac{\partial^{n}\widetilde{\rho}_N({\bf r})}{\partial N^n}
\bigg|_{N=\langle N\rangle}
\!=\;\;
\frac{\partial^{n}\!\rho_{\langle N\rangle}({\bf r})}{\partial 
\langle N\rangle^{n}}\bigg|_{\langle N\rangle=N}
\end{equation}
which indeed seems plausible, but is surely 
difficult to justify in any rigorous way.
We thus generate the following expansion
\begin{align}\label{Gonzalez5}
&\rho_{\langle N\rangle}({\bf r})
=
\rho({\bf r})
\,-\, 
\frac{1}{2}
\left\langle \left( N \!-\!\langle N\rangle\,\right)^{2}
\right\rangle
\frac{\partial^{2}\!
\rho_{\langle N\rangle}({\bf r})}{\partial 
\langle N\rangle^{2}}\bigg|_{\langle N\rangle=N}
\notag\\
&\;\;\;\;-\,
\frac{1}{6}
\left\langle \left( N \!-\!\langle N\rangle\,\right)^{3}
\right\rangle
\frac{\partial^{3}\!
\rho_{\langle N\rangle}({\bf r})}{\partial \langle N\rangle^3}\bigg|_{\langle N\rangle=N}
\,-\;\cdots\;
,  
\end{align}
Equation \eqref{Gonzalez5} forms the basis for an iterative back-substitution scheme (termed `formal inversion' in the original papers 
of Gonz\'alez {\it et al.}). 
Recursive substitution of the entire right-hand side 
of \eqref{Gonzalez5} into the partial derivatives 
then produces the main result, equation \eqref{eq_evans_expansion}, given in the main text.

It is quite remarkable to us that this recursive scheme successfully generates 
the leading terms of an asymptotic expansion, despite not 
having an obvious connection to the asymptotic evaluation 
of any integral. 
The deeper reasons for why our systematic expansion 
\eqref{final_expansion_density} coincides with the 
Gonz\'alez expression, at least for the terms we could check, 
will require further investigation and at present remains 
obscure. 
As a final comment we mention that the Gonz\'alez derivation 
does not offer much insight into when and why the 
expansion may fail. As far as we can see this information 
can only be gained by analyzing the grand-canonical to 
canonical transform in the complex fugacity plane.

\bibliographystyle{apsrev4-2} 
\bibliography{bibliography}

\end{document}